\documentclass[prl,twocolumn]{revtex4-1}
\usepackage{graphicx}
\usepackage{amsmath}
 \usepackage{amssymb}
\usepackage{amsfonts}
\usepackage{upgreek}
\usepackage{comment}
\usepackage{color}
\usepackage{hyperref}
\usepackage{soul}
\hypersetup{colorlinks=true, linkcolor=blue, citecolor=blue, urlcolor=blue}

\begin{document}

\title{A Superradiant Topological Peierls Insulator inside an Optical Cavity}
 
 \author{Farokh Mivehvar, Helmut Ritsch, and Francesco Piazza}
 \email[Corresponding author: ]{francesco.piazza@uibk.ac.at}
\affiliation{\small Institut f\"ur Theoretische Physik, Universit{\"a}t Innsbruck, A-6020~Innsbruck, Austria}

\begin{abstract}
We consider a spinless ultracold Fermi gas tightly trapped along the axis of an optical resonator and transversely illuminated by a laser closely tuned to a resonator mode. At a certain threshold pump intensity the homogeneous gas density breaks a $\mathbf{Z}_2$ symmetry towards a spatially periodic order, which collectively scatters pump photons into the cavity. We show that this known self-ordering transition also occurs for low field seeking fermionic particles when the laser light is blue-detuned to an atomic transition. The emergent superradiant optical lattice in this case is homopolar and possesses two distinct dimerizations. Depending on the spontaneously chosen dimerization the resulting Bloch bands can have a non-trivial topological structure characterized by a non-vanishing Zak phase.  In the case the Fermi momentum is close to half the cavity-mode wavenumber, a Peierls-like instability here creates a topoloical insulator with a gap at the Fermi surface, which hosts a pair of edge states. The topological features of the system can be non-destructively observed via the cavity output: the Zak phase of the bulk coincides with the relative phase between laser and cavity field, while the fingerprint of edge states can be observed as additional broadening in a well defined frequency window of the cavity spectrum.

\end{abstract}

\maketitle

\emph{Introduction.}---The experimental progress in coupling ultracold quantum gases to the electromagnetic field of high-Q cavities~\cite{vuletic_2003,courteille_2007, reichel_2007,st_kurn_2007,eth_2010,barrett_2012,hemmerich_SR_2014,lev_multimode_exp_2015} has opened a new avenue for creating and exploring novel many-body collective phenomena in
the framework of cavity quantum electrodynamics (QED)~\cite{cavity_rmp,maschler_2008,gopal_2009,gopal_glass_2011,strack_2011,caballero_2015,mazzucchi2016quantum,torggler2016quantum}. 
One hallmark effect of the coupled atom-field dynamics is self-ordering, where the atoms spontaneously break the translational symmetry and form a spatial pattern, which maximizes collective (i.e., superradiant) scattering of the pump photons into the cavity~\cite{vuletic_2003,barrett_2012,eth_2010,hemmerich_SR_2014,lev_supermode,ritsch_2002}. The interplay between light-induced long-range interactions, the quantum statistics of the particles~\cite{domokos_2008,piazza_bose,lode2016fragmented,keeling_2014,zhai_2014,piazza_fermi,piazza_QKE,sandner_fermi}
and short-range interatomic interactions~\cite{morigi_2013,hofstetter13,bakhtiari_2015, zhai_mottcav_2016,dogra2016phase,morigi_competing_2016,gelhausen2016quantum}
gives rise to a wealth of intriguing phenomena. Corresponding experiments have become successful quantum simulators for Bose-Hubbard models with infinitely long-range interactions demonstrating the Dicke superradiant phase transition as well as a supersolid phase~\cite{hemmerich_mottcav_2015,landig2016quantum,leonard2016supersolid}.

Recent theoretical developments have highlighted a further possibility to exploit cavity fields to generate artificial
spin-orbit~\cite{mivehvar2014synthetic,sr_soc_yi_2014,cavity_soc_Gosh_2014,cavity_soc_Han_2014,sr_soc_fermi_guo_2015}
or dynamical gauge fields~\cite{kollath2016ultracold,sheikhan2016cavity,dicke_spinHall_2015, zheng2016superradiance,keeling_meissner_cav_2016}, exploiting Raman processes involving a cavity mode to induce internal transitions between two atomic ground-state sublevels (pseudospins)~\cite{sr_soc_yi_2014,sr_soc_fermi_guo_2015} or
tunneling between two sites of a pre-existing lattice~\cite{kollath2016ultracold,sheikhan2016cavity, zheng2016superradiance}.
As a consequence, self-organized phases are predicted to become topological when the artificial spin-orbit coupling or the
gauge field is mediated by the superradiant cavity light~\cite{sr_soc_fermi_guo_2015,kollath2016ultracold, sheikhan2016cavity}.

\begin{figure}[b]
\includegraphics[width=\columnwidth]{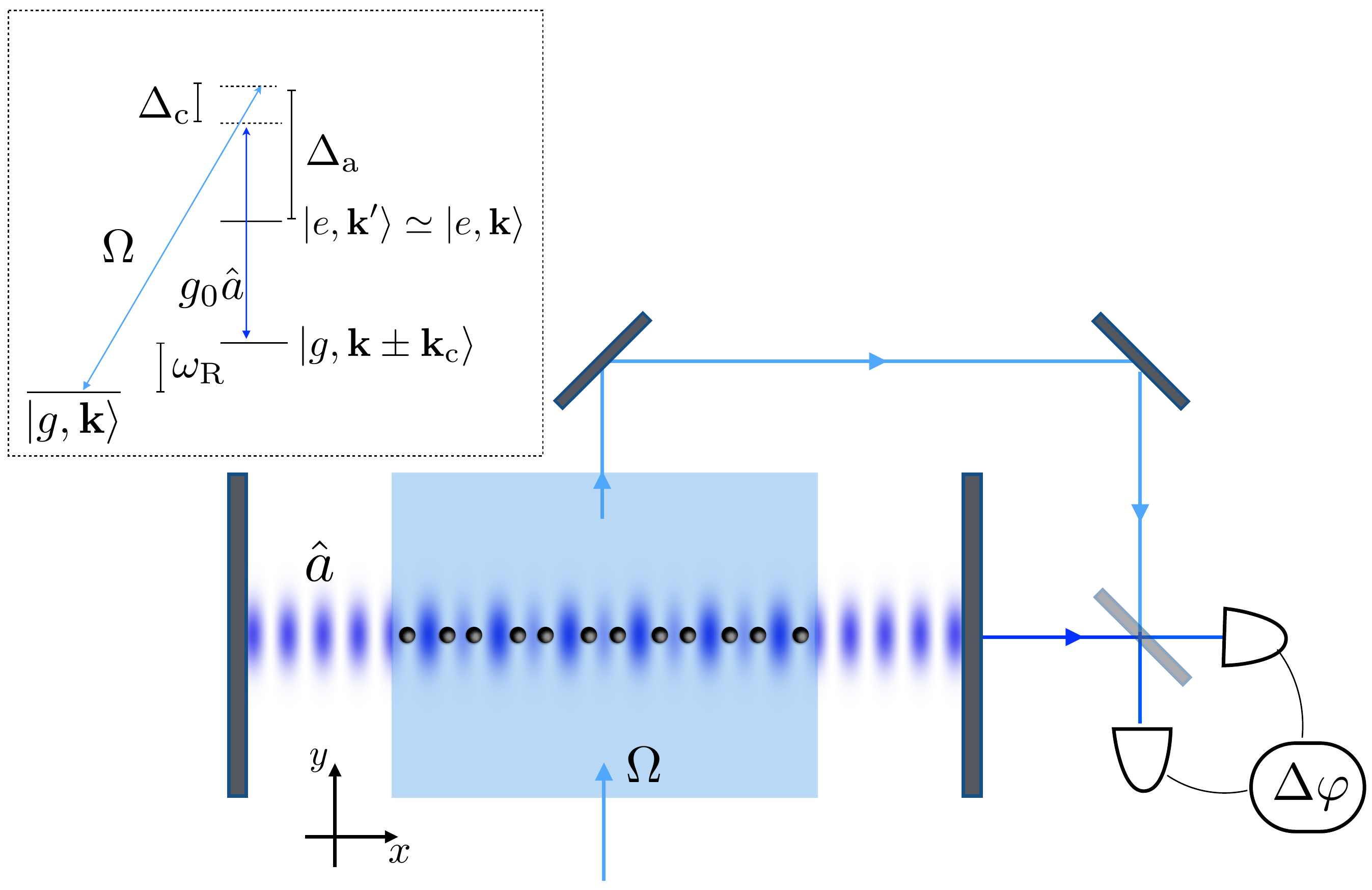}
\caption{Schematic view of fermonic atoms trapped in a 1D elongated tube along the axis of an optical resonator and driven by a  transverse blue-detuned ($\Delta_{\rm a}>0$ with respect to an atomic transition $g\leftrightarrow e$) laser  with Rabi frequency $\Omega$.  $\Delta_c$ denotes detuning with respect to a standing-wave cavity mode. The atoms are low field seekers trapped at the light intensity minima generated by the interference of the cavity field and plane-wave transverse pump laser. The relative phase $\Delta\varphi$ of cavity output and pump laser corresponds to the Zak phase of the lattice bands in real time (see Fig.~\ref{fig:symm}). The inset shows the coupling of momentum states via pump and cavity fields.}
\label{fig:setup}
\end{figure}
In this Letter, we present a configuration leading to topologically non-trivial self-ordered phases for spin-polarized fermionic atoms in one dimension without artificial spin-orbit coupling or gauge fields.  We consider the same experimental setup exploited to observe the superradiant self ordering transition of bosons~\cite{vuletic_2003,barrett_2012,eth_2010,hemmerich_SR_2014}, involving transversely laser-driven atoms coupled to a single mode of an optical resonator in the dispersive regime (see Fig.~\ref{fig:setup}). The motion of the atoms transverse to the cavity axis is frozen by a cigar-shaped dipole trap~\cite{landig2016quantum}.
The superradiant lattice formed from the interference between the laser and the cavity field is dimerized, i.e., the unit-cell contains two lattice sites. During self ordering the particles choose between the two possible dimerizations in a spontaneous $\mathbf{Z}_2$-symmetry breaking process. The nature of the dimerized superradiant lattice qualitatively depends on the sign of the laser-atom detuning $\Delta_a$. For the conventional red detuning $\Delta_a<0$, dimerization is heteropolar (i.e., there is a finite energy offset between the two sites in the unit cell), while for blue detuning, on which we focus in the following, $\Delta_a>0$ it is homopolar (zero offset)~\cite{piazza_blue}.
The dimerized-lattice bands can have a non-trivial topological structure characterized by their Zak phase $\varphi_{\rm Zak}$, which is the Berry's phase picked
up by moving adiabatically through the entire first Brillouin zone~\cite{zak_1989}.
In the homopolar (blue-detuned) case, $\varphi_{\rm
 Zak}$ is $\mathbf{Z}_2$ quantized and can be either $0$ or $\pi$ 
 depending on the (in our case spontaneously chosen) dimerization. 

Our model system shares several features with the Su-Schrieffer-Heeger (SSH)~\cite{ssh_1979} and the Holstein~\cite{holstein_1959} models, describing electrons coupled to lattice phonons. In our case, the role of the many phonon modes is played by a single global cavity mode.
In particular, if the Fermi momentum is close to half the cavity-mode wave-vector, a Peierls-like instability creates a superradiant dimerized lattice which opens a gap at the Fermi surface~\cite{piazza_fermi}. In this case for one of the two possible dimerizations, the system becomes a topological insulator~\cite{hasan_kane,ryu2010topological} with chiral symmetry, hosting a pair of edge states within the gap in the finite system.

We also demonstrate how the signatures of the non-trivial topology
in our open system can be non-destructively read out in real time simply by monitoring the cavity output. The Zak
phase directly appears as the relative phase $\Delta\varphi$ between the laser and cavity field,
which is an easily accessible quantity~\cite{eth_jumps}. In addition, in a finite system the edge states can be detected in the cavity spectrum since their presence introduces additional broadening in a specific well-defined frequency window.

\begin{figure}[t]
\includegraphics[width=\columnwidth]{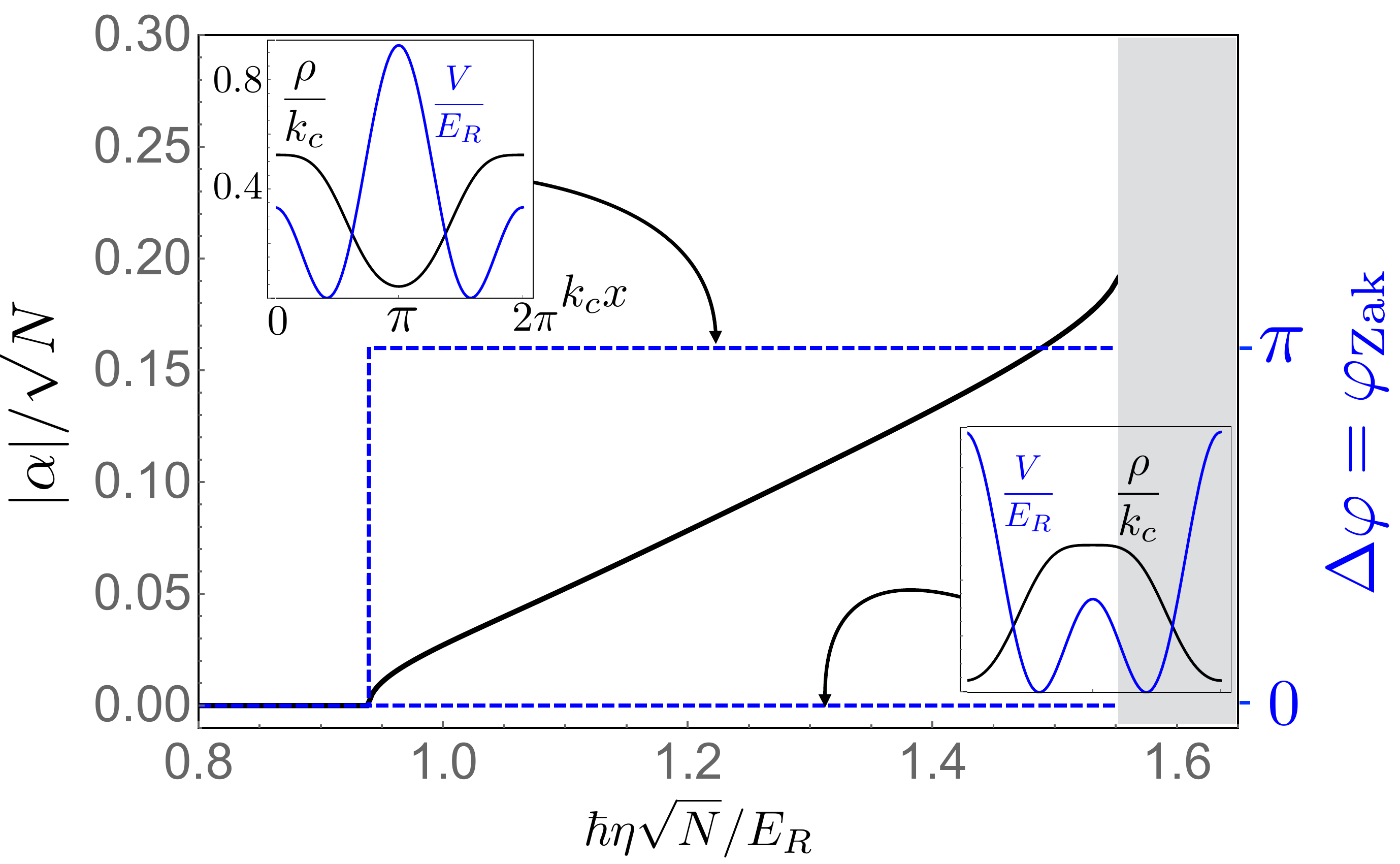}
\caption{Spontaneous $\mathbf{Z}_2$-symmetry breaking at the superradiant self-ordering transition in the configuration of  Fig.~\ref{fig:setup} seen via the modulus of coherent cavity field amplitude (solid line) as  function of the pump strength $\eta$. The blue dashed line shows the cavity phase $\Delta\varphi$ relative to the driving laser, for which we have either $\Delta\varphi=0$ or $\Delta\varphi=\pi$ depending on the lattice dimerization. The insets show the corresponding atomic densities (black solid line) and lattice potential (rescaled by $0.5$, blue solid line) within a unit cell, for $\hbar\eta\sqrt{N}=1.55 E_R$.  $\Delta\varphi$  coincides with the Zak phase $\varphi_{\rm Zak}$ of the lowest Bloch band. The gray-shaded area denotes an unstable region of the system (see Fig.~\ref{fig:ph_diag}). Results are obtained for spin-polarized fermions with parameters: $k_F=k_c/2$, $k_B T=0.01 E_R$, $\hbar\Delta_c=12 E_R$, $U_0N=32.35 E_R$.}
\label{fig:symm}
\end{figure}
\emph{Model and approach.}---Consider laser-driven ultracold spin-polarized 
fermions in a standing-wave resonator as illustrated in Fig.~\ref{fig:setup}. 
The system is effectively described by the Hamiltonian \cite{maschler_2008},
\begin{align}
\hat{H}
=\int dx \hat{\psi}^{\dag} \Big[
&-\frac{\hbar^2}{2m}\frac{d^2}{dx^2} 
+\hbar\eta\cos(k_cx)\left(\hat{a}e^{-i\varphi_\Omega}+\text{H.c.}\right)\nonumber\\ 
&+U_0\cos^2(k_c x)\hat{a}^\dagger\hat{a}\Big]
\hat{\psi}
-\hbar\Delta_{\rm c}\hat{a}^{\dag}\hat{a},
\label{eq:H}
\end{align}
where the atomic excited state has been adiabatically eliminated (dispersive regime of large $\Delta_a$) and
the atomic motion is restricted to one dimension along the cavity axis $x$. We assume plane-wave laser illumination and a standing-wave cavity mode proportional to $\cos(k_c x)$ with wavenumber $k_c$. 
In Eq.~\eqref{eq:H} we introduced the optical potential depth per photon
$U_0\equiv\hbar g_0^2/\Delta_a$ equivalent to the cavity frequency shift per atom, the effective pump strength 
$\eta\equiv g_0|\Omega|/\Delta_a$, the laser phase $\varphi_\Omega$ and the cavity detuning with respect to the pump frequency $\Delta_c$.
Here, $g_0$ and $\Omega$ are the atom-cavity coupling and the single-photon Rabi frequency, respectively, and H.c.~stands for the Hermitian conjugate. The recoil energy $E_R\equiv\hbar^2k_c^2/2m$ and cavity wavenumber $k_c$ are the natural units
of energy and momentum here. 
Finally, $\hat{a}$ is the cavity photon bosonic annihilation operator  and $\hat{\psi}(x)$ is the atomic fermionic field
operator. 

The $\mathbf{Z}_2$-symmetry transformation $x\to
x\pm\pi/k_c$, $\hat{a}\to -\hat{a}$, $\hat{a}^\dag\to -\hat{a}^\dag$
leaves the Hamiltonian \eqref{eq:H} invariant. This symmetry is
spontaneously broken at the self-ordering transition where the cavity mode develops a finite coherent amplitude
$\alpha=\langle\hat{a}\rangle=|\alpha|\exp(i\varphi_\alpha)$ and its
phase is locked to the laser phase $\varphi_\Omega$ by a
Josephson-like energy term
$E_J(\alpha)\propto -N\eta^2\rho|\alpha|^2\cos^2(\Delta\varphi)$,
with $\rho=N/L$ the average atomic density~\cite{piazza_bose}. $E_J$ fixes the relative phase
$\Delta\varphi=\varphi_\alpha-\varphi_\Omega$ to being either $0$ or
$\pi$ (see Fig.~\ref{fig:symm}). 
In the thermodynamic limit: $N,L\to\infty$, $\rho=$const., the
superradiant phase is exactly described by a mean-field approach
where only the coherent field component $\alpha$ is
retained and satisfies the following self-consistent equations \cite{piazza_bose}:
\begin{align}
\alpha=\frac{1}{\Delta_c}\int\!\!  dx \rho(x)\frac{\partial V_{\rm
    sl}(x)}{\partial\alpha^*},\; \int\!\! dx\rho(x)=N,
\label{eq:mf}
\end{align}
with the superradiant lattice potential
\begin{align}
V_{\rm  sl}(x)=U_0|\alpha|^2\cos^2(k_cx)+2\hbar\eta |\alpha|\cos(\Delta\varphi)\cos(k_cx),
\label{eq:srlattice}
\end{align}
and the atomic density
\begin{align}
\rho(x)=\langle\hat{\psi}^\dag(x)\hat{\psi}(x)\rangle
=\sum_{n}\int_{\text{BZ}}\frac{dk}{2\pi}n_F\left(\epsilon_n(k)\right) |\psi_{n,k}(x)|^2.
\label{eq:den}
\end{align}
Here, $n_F(\epsilon)=[1+\exp(\beta(\epsilon-\mu))]^{-1}$ 
is the Fermi-Dirac distribution, and $\epsilon_n(k)$ and $\psi_{n,k}(x)$ are, respectively, 
the Bloch eigenenergies and eigenstates of the single-particle
Hamiltonian with the superradiant lattice potential $V_{\rm sl}(x)$,
whose periodicity $2\pi/k_c$ defines the Brillouin zone
$\text{BZ}=[-k_c/2,k_c/2]$. The
chemical potential is fixed by the second equation in~\eqref{eq:mf}. We stress that the dependence of
$\rho(x)$ on $\alpha$  through the dynamical lattice $V_{\rm sl}(x)$ requires a self-consistent solution for $\alpha$ and $\mu$ in Eq.~\eqref{eq:mf}.

Dealing with an open system one should in principle include photon losses at rate $\kappa$ out of the cavity~\cite{cavity_rmp}. However, for  very good resonators the induced heating rates of the Fermi gas are ineffective up to long timescales proportional to the atom number $N$, as shown in Ref.~\cite{piazza_QKE}. Within the superradiant phase the losses introduce damping of the coherent field amplitude, which follows the momentary atomic distribution with some small phase shift~\cite{eth_jumps,eth_non_eq,hemmerich_dyn_2014}. Even though the superradiant self-ordering  transition becomes dissipative the $\mathbf{Z}_2$-nature of the symmetry breaking is not affected.

\emph{Emergent dimerized superradiant lattice.}---The solution of
Eq.~(\ref{eq:mf}) is shown in Fig.~\ref{fig:symm}. Above a critical
pump-strength (see also Fig.~\ref{fig:ph_diag} below) the coherent
part of the cavity field  $\alpha$ grows monotonically with its relative phase $\Delta\varphi$ locked at either 0 or $\pi$, breaking the $\mathbf{Z}_2$ symmetry. The two cases correspond
to two possible lattice dimerizations, as shown by Eq.~(\ref{eq:srlattice}) and in the
inset of Fig.~\ref{fig:symm} within one unit-cell. 

\begin{figure}[t]
\includegraphics[width=\columnwidth]{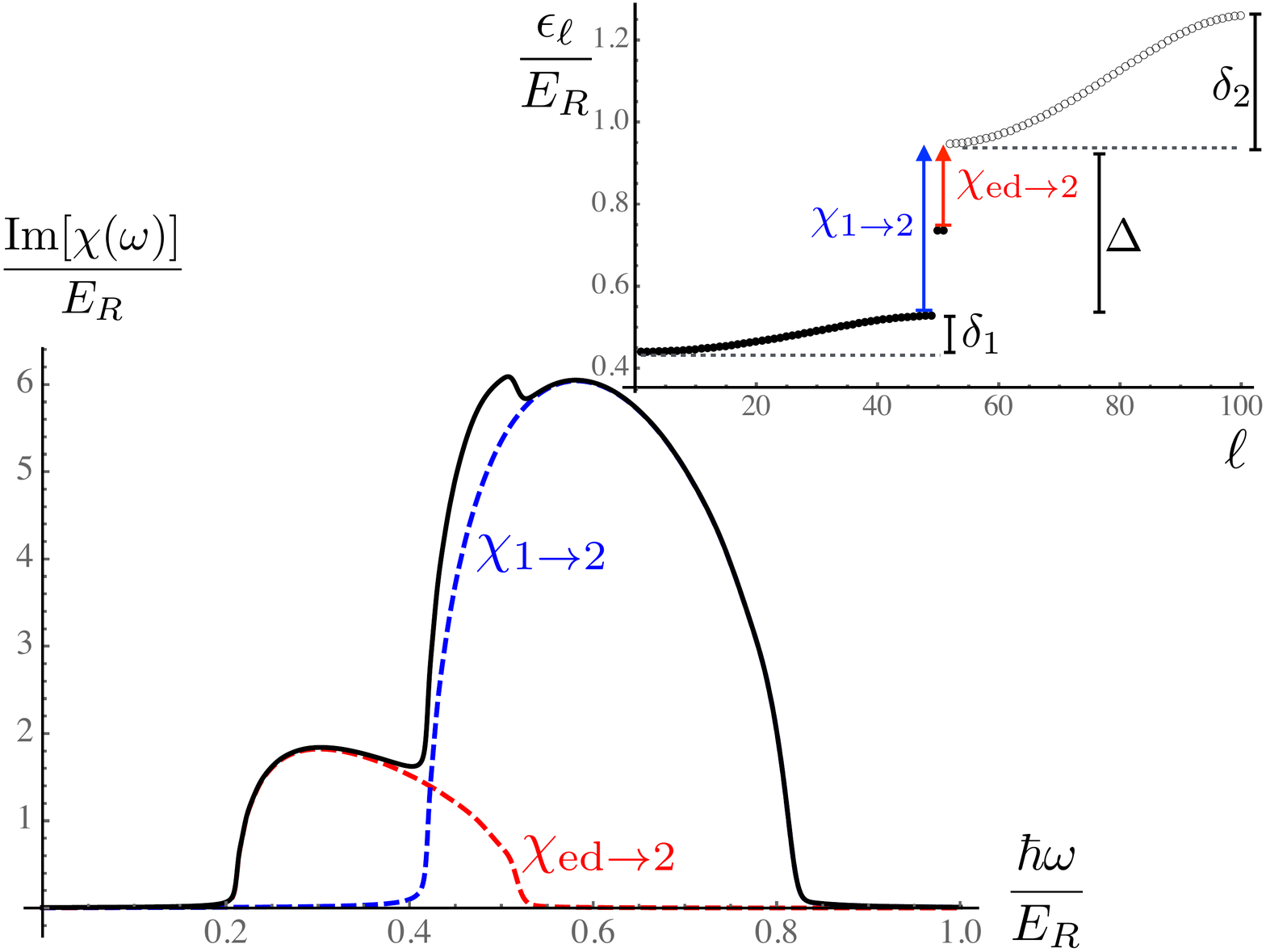}
\caption{Signature of the edge states in the absorption spectrum
of the SRTI phase with respect to the
  propagation of cavity photons. The imaginary part of the
  polarization [Eq.~\eqref{eq:pol}] as a function of frequency is shown for a finite system
encompassing $50$ unit cells at half filling.
The total absorption $\chi$ (black solid
line) has two contributions: $\chi_{\rm ed\to 2}$ (red dashed line) is
due to the presence of edge states, while $\chi_{1\to 2}$ (blue dashed line) is present also without edge states. 
 The inset shows the atomic energy spectrum, where filled (empty) circles indicate occupied (empty) states. The two isolated
eigenenergies in the mid-gap correspond to the two localized edge states.
 The parameters are the same as in
Fig.~\ref{fig:symm} with $\hbar\eta\sqrt{N}=1.55E_R$.
}
\label{fig:polarization}
\end{figure}

The one dimensional bands can be characterized by the Zak phase~\cite{zak_1989,resta_rmp}, 
\begin{align}
\varphi_{\rm Zak}^{(n)}=k_c\int_\text{BZ}\!dk\int_0^{\frac{2\pi}{k_c}}\!\!\!dx\;u_{n,k}^*(x)\frac{\partial u_{n,k}(x)}{\partial
k}\!\overset{n=1}{=}\Delta\varphi,
\label{eq:zak}
\end{align}
where $u_{n,k}(x)=\exp(-ikx) \psi_{n,k}(x)$ is the periodic part of the Bloch wave-function. 
The Zak phase is in general not quantized and it can take any value between
zero and $2\pi$. That said, in the presence of chiral symmetry (as for the blue-detuned, homopolar lattice considered here)
it is $\mathbf{Z}_2$ quantized and can assume solely the values $0$ or $\pi$. The Zak phase $\varphi_{\rm Zak}$ of the 
lowest Bloch band $n=1$ coincides with the relative
phase of the cavity $\Delta\varphi$, as indicated in the last equality of Eq.~\eqref{eq:zak}, in case of a mirror symmetric choice for the unit cell (see the inset of Fig.~\ref{fig:symm}).
The mirror symmetric choice of the unit cell, though not being a unique choice~
\cite{vanderbilt_1993,bloch_zak}, corresponds to the Wigner-Seitz cell; and
the Zak phase can then be interpreted as a topological invariant. 
The non-vanishing Zak phase $\varphi_{\rm Zak}=\pi$ signals indeed a nontrivial topological band,
which hosts localized zero-dimensional edge states in a finite system.
This is illustrated in the inset of Fig.~\ref{fig:polarization}, showing the low-lying 
energy spectrum of a finite lattice consisting of 50 unit cells of the topologically
non-trivial dimerization
. A smooth polynomial wall potential 
$V_{\rm{wall}}(x)=V_0(x^2+x^4+x^6)$ is added to both ends of the finite lattice,
with $V_0/E_R=150$. 
The two isolated eigen-energies in the mid-gap correspond to the two localized edge states.
They are quite robust with respect to variation of the wall potential in the range 
$V_0/E_R\sim1-10^4$. However, they lie around mid-gap only for $V_0/E_R\sim10^2$.

Measuring topological invariants directly is often challenging and requires non-local probes, 
as for instance the Bloch-Ramsey interferometric technique employed to measure the Zak phase with ultracold atoms in a
super-lattice~\cite{bloch_zak}. In our setup where the lattice (\ref{eq:srlattice}) is formed through the interference between the laser and the superradiant cavity field, $\varphi_{\rm Zak}$ can be continuously and non-destructively read out
as the relative phase $\Delta\varphi$ between the pump and cavity fields, as schematically depicted 
in Fig.~\ref{fig:setup}.

\begin{figure}[t]
\includegraphics[width=\columnwidth]{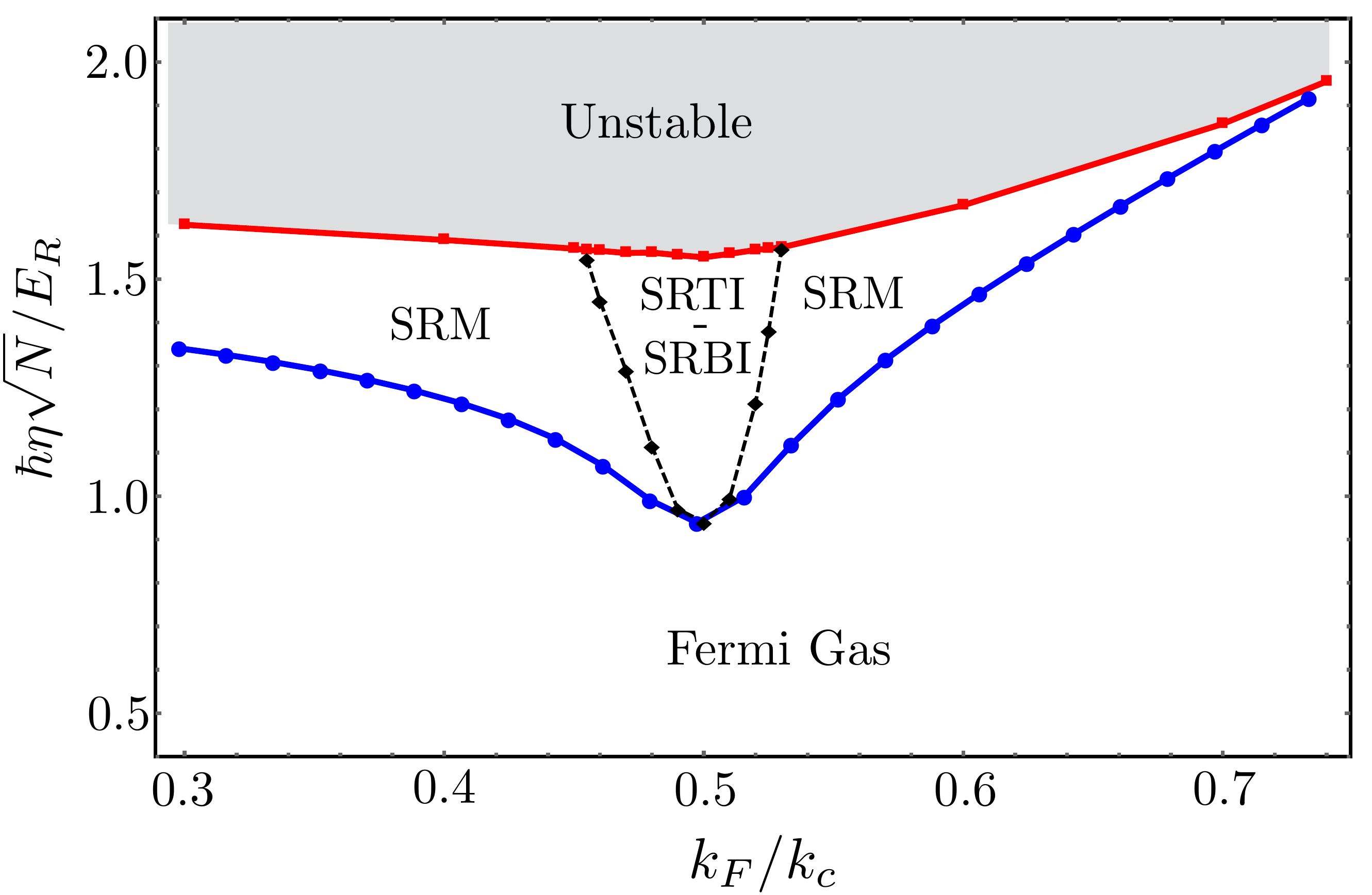}
\caption{Phase diagram in the $\eta-k_F$ plane. For strong enough pump $\eta$ (above the blue line) the system self-orders into a superradiant pattern. In the self-ordered phase, the system is either in a SRM phase or an insulator state.
The superradiant insulating phase, in turn, is either a SRBI or SRTI, characterized by their Zak phases (see Fig.~\ref{fig:symm}).
At half filling $k_F=k_c/2$ we have a direct phase transition between the uniform Fermi gas and the SRBI/SRTI. Otherwise this phase transition occurs only through the SRM phase. For even stronger drive (above red line) the self-ordered phase becomes unstable. The parameters are the same as in Fig.~\ref{fig:symm}.}
\label{fig:ph_diag}
\end{figure}
\emph{Topological insulator.}---Figure~\ref{fig:ph_diag} shows the phase diagram in the parameter space of 
the effective pump strength $\hbar\eta\sqrt{N}/E_R$ versus the Fermi momentum $k_F/k_c$.
For densities around half-filling $k_F=k_c/2$ with $k_F=\pi \rho$, the superradiant phase corresponds to an
insulator, while sufficently away from this point the system is found in a superradiant metallic (SRM) phase. The dashed lines in Fig.~\ref{fig:ph_diag} indicate the continuous crossover
from the metallic to insulating phase. 
The transition to an insulator within a dimerized lattice at
half-filling follows the same mechanism underlying the Peierls
instability in electron-phonon models \cite{peierls_1955,ssh_1979},
but the phonons here are substituted by photons in the cavity mode.
The critical pump strength for the superradiant transition vanishes at $T=0$, as noted already for red-detuned superradiant lattices \cite{keeling_2014,piazza_fermi,zhai_2014}. 
The superradiant insulating phase has a band gap $\Delta$ proportional to the absolute value
of the cavity-field amplitude $|\alpha|$.  The superradiant insulator phase can be either a (topologically trivial) band insulator (SRBI) for $\Delta\varphi=0$ or a topological insulator (SRTI) for $\Delta\varphi=\pi$. For the latter case the lattice bands host a pair of edge states in the finite system, as discussed above.
The SRTI possesses chiral (or sublattice) symmetry~\cite{hatsugai_2002} and belongs to the AIII class introduced in Ref.~\cite{ryu2010topological}. 
In the tight-binding limit, particle-hole and time-reversal symmetries are additionally present,
putting the SRTI in the BDI class of Ref.~\cite{ryu2010topological}, i.e., in the same class as the SSH model.

Increasing the pump amplitude $\eta$ further, the system reaches a second transition point where no stable solutions of the
Eqs.~\eqref{eq:mf} can be found anymore as indicated by the gray-shaded areas in Figs.~\ref{fig:symm} and~\ref{fig:ph_diag}. This instability is caused by the competition
between the $\propto\cos(k_c x)$ and the $\propto\cos^2(k_c x)$ contribution to the superradiant lattice potential in
Eq.~\eqref{eq:srlattice} and is characteristic of the blue-detuned homopolar lattice.
As shown in Ref.~\cite{piazza_blue}, this instability can correspond to the onset of limit-cycle and even chaotic behaviour.

\emph{Detecting edge states.---}As described above, in a finite system the SRTI phase possesses a pair of topological edge states within the band gap, which can be directly observed in the spectral properties of the
cavity output. This can be verified by computing the optical polarzability $\chi(\omega)$ of the SRTI in the finite system.  $\chi(\omega)$ corresponds to the dynamic response function of the medium with respect to density perturbations induced by cavity photons and reads~\footnote{See Supplemental Material},
\begin{align} 
\chi(\omega)=\sum_{\ell,\ell'}\frac{n_F(\epsilon_\ell)-n_F(\epsilon_{\ell '})}
{\hbar\omega+\epsilon_\ell-\epsilon_{\ell '} +i0^+}\bigg|\int\!\! dx\psi_\ell^* 
\frac{\partial V_{\rm sl}(x)}{\partial\alpha^*}\psi_{\ell '}\bigg|^2,
\label{eq:pol}
\end{align}
where $\epsilon_\ell$ and $\psi_\ell(x)$ are the eigenenergies and eigensates
of the single-particle Hamiltonian with potential $V_{\rm sl}(x)+V_{\rm wall}(x)$. 
The real (imaginary) part of $\chi$ corresponds to the atomic dispersion (absorption) with respect to the propagation of cavity
photons. 
%


Using the identity
$\mathrm{Im}[1/(\hbar\omega+\epsilon_{\ell}
  -\epsilon_{\ell'} +i0^+)]=-\pi\delta(\hbar\omega+\epsilon_{\ell}
  -\epsilon_{\ell'})$, one can see that the medium absorbs photons only if  energy is conserved in the transition between two eigenstates (one of which must be occupied) having
  a finite matrix element with respect to the operator $\partial
  V_{\rm sl}/\partial\alpha^*$. In the absence of edge states, the Fermi medium absorbs cavity photons
only for frequencies $\hbar\omega\geqslant\Delta$, where $\Delta$ is the energy gap between
low-lying filled valence states and upper empty conduction states (see the inset 
in Fig.~\ref{fig:polarization}) and is assumed to be much larger than the temperature, $k_BT\ll \Delta$.
On the other hand, when edge states are present in the middle of the gap the medium can 
absorb photons with even lower energy: $\hbar\omega\geqslant\Delta/2$. This holds as long as the
matrix element is non-vanishing. By increasing the length of the system $L$, the edge-state
contribution $\chi_{\rm ed\to 2}(\omega)$ to the total response
function $\chi(\omega)$ decreases like $1/L$ in the thermodynamic limit $N/L=$const., where also $U_0\propto 1/L$ and $\eta\propto 1/\sqrt{L}$. 
The edge states indeed only locally contribute to absorption.

Figure~\ref{fig:polarization} shows the imaginary part of $\chi(\omega)$.
The blue (red) dashed curve indicates $\chi_{1\to 2}$ ($\chi_{\rm ed\to 2}$), 
the contribution of transitions from filled
lowest valence (edge) states into upper empty conduction states.
It is evident that the presence of edge
states in the SRTI phase drastically modifies the 
polarization function and opens an
additional absorption channel in the well defined frequency range
$\Delta/2\leqslant\hbar\omega\leqslant \Delta/2+\delta_2$, with
$\delta_2$ being the conduction bandwidth. We note
that for $\delta_2<\Delta/2$ this additional absorption window becomes
even fully separated from the one related to transitions between valence
and conduction states.
This provides experimental means to detect edge states via a non-destructive measurement, 
  since the absorption $\mathrm{Im}[\chi(\omega)]$ influences the
  broadening of the cavity resonance and can be for instance extracted
  from the width of the peak of the incoherent fluorescence spectrum
  or probe-transmission spectrum \cite{piazza_QKE}. 

\emph{Conclusions.---} We introduced and characterized a simple configuration of fermionic atoms trapped along the axis of an optical cavity to implement a genuine topological self-organized superradiant state. The proposed setup is already experimentally available by a simple change of the trapping laser frequency towards blue atom-pump detuning without any need of creating artificial spin-orbit coupling or gauge fields. The interference of pump and cavity field here automatically creates a dimerized homopolar lattice so that
fermionic atoms self-order into a topological insulator close to half filling at low enough temperature. Remarkably, the configuration features built-in non-destructive monitoring tool via the cavity output photons, which allows to directly probe the topological properties of
the system like the Zak phase and the existence of the edge states. 

\subsection{Acknowledgments}
We thank Johannes Lang for fruitful discussions.
FP acknowledges support by the APART fellowship of the Austrian Academy of Sciences. FM and HR are supported by the Austrian Science Fund project I1697-N27.

\bibliography{mybib}

\begin{thebibliography}{62}%
\makeatletter
\providecommand \@ifxundefined [1]{%
 \@ifx{#1\undefined}
}%
\providecommand \@ifnum [1]{%
 \ifnum #1\expandafter \@firstoftwo
 \else \expandafter \@secondoftwo
 \fi
}%
\providecommand \@ifx [1]{%
 \ifx #1\expandafter \@firstoftwo
 \else \expandafter \@secondoftwo
 \fi
}%
\providecommand \natexlab [1]{#1}%
\providecommand \enquote  [1]{``#1''}%
\providecommand \bibnamefont  [1]{#1}%
\providecommand \bibfnamefont [1]{#1}%
\providecommand \citenamefont [1]{#1}%
\providecommand \href@noop [0]{\@secondoftwo}%
\providecommand \href [0]{\begingroup \@sanitize@url \@href}%
\providecommand \@href[1]{\@@startlink{#1}\@@href}%
\providecommand \@@href[1]{\endgroup#1\@@endlink}%
\providecommand \@sanitize@url [0]{\catcode `\\12\catcode `\$12\catcode
  `\&12\catcode `\#12\catcode `\^12\catcode `\_12\catcode `\%12\relax}%
\providecommand \@@startlink[1]{}%
\providecommand \@@endlink[0]{}%
\providecommand \url  [0]{\begingroup\@sanitize@url \@url }%
\providecommand \@url [1]{\endgroup\@href {#1}{\urlprefix }}%
\providecommand \urlprefix  [0]{URL }%
\providecommand \Eprint [0]{\href }%
\providecommand \doibase [0]{http://dx.doi.org/}%
\providecommand \selectlanguage [0]{\@gobble}%
\providecommand \bibinfo  [0]{\@secondoftwo}%
\providecommand \bibfield  [0]{\@secondoftwo}%
\providecommand \translation [1]{[#1]}%
\providecommand \BibitemOpen [0]{}%
\providecommand \bibitemStop [0]{}%
\providecommand \bibitemNoStop [0]{.\EOS\space}%
\providecommand \EOS [0]{\spacefactor3000\relax}%
\providecommand \BibitemShut  [1]{\csname bibitem#1\endcsname}%
\let\auto@bib@innerbib\@empty
\bibitem [{\citenamefont {Black}\ \emph {et~al.}(2003)\citenamefont {Black},
  \citenamefont {Chan},\ and\ \citenamefont {Vuleti\ifmmode~\acute{c}\else
  \'{c}\fi{}}}]{vuletic_2003}%
  \BibitemOpen
  \bibfield  {author} {\bibinfo {author} {\bibfnamefont {A.~T.}\ \bibnamefont
  {Black}}, \bibinfo {author} {\bibfnamefont {H.~W.}\ \bibnamefont {Chan}}, \
  and\ \bibinfo {author} {\bibfnamefont {V.}~\bibnamefont
  {Vuleti\ifmmode~\acute{c}\else \'{c}\fi{}}},\ }\href {\doibase
  10.1103/PhysRevLett.91.203001} {\bibfield  {journal} {\bibinfo  {journal}
  {Phys. Rev. Lett.}\ }\textbf {\bibinfo {volume} {91}},\ \bibinfo {pages}
  {203001} (\bibinfo {year} {2003})}\BibitemShut {NoStop}%
\bibitem [{\citenamefont {Slama}\ \emph {et~al.}(2007)\citenamefont {Slama},
  \citenamefont {Bux}, \citenamefont {Krenz}, \citenamefont {Zimmermann},\ and\
  \citenamefont {Courteille}}]{courteille_2007}%
  \BibitemOpen
  \bibfield  {author} {\bibinfo {author} {\bibfnamefont {S.}~\bibnamefont
  {Slama}}, \bibinfo {author} {\bibfnamefont {S.}~\bibnamefont {Bux}}, \bibinfo
  {author} {\bibfnamefont {G.}~\bibnamefont {Krenz}}, \bibinfo {author}
  {\bibfnamefont {C.}~\bibnamefont {Zimmermann}}, \ and\ \bibinfo {author}
  {\bibfnamefont {P.~W.}\ \bibnamefont {Courteille}},\ }\href {\doibase
  10.1103/PhysRevLett.98.053603} {\bibfield  {journal} {\bibinfo  {journal}
  {Phys. Rev. Lett.}\ }\textbf {\bibinfo {volume} {98}},\ \bibinfo {pages}
  {053603} (\bibinfo {year} {2007})}\BibitemShut {NoStop}%
\bibitem [{\citenamefont {Colombe}\ \emph {et~al.}(2007)\citenamefont
  {Colombe}, \citenamefont {Steinmetz}, \citenamefont {Dubois}, \citenamefont
  {Linke}, \citenamefont {Hunger},\ and\ \citenamefont
  {Reichel}}]{reichel_2007}%
  \BibitemOpen
  \bibfield  {author} {\bibinfo {author} {\bibfnamefont {Y.}~\bibnamefont
  {Colombe}}, \bibinfo {author} {\bibfnamefont {T.}~\bibnamefont {Steinmetz}},
  \bibinfo {author} {\bibfnamefont {G.}~\bibnamefont {Dubois}}, \bibinfo
  {author} {\bibfnamefont {F.}~\bibnamefont {Linke}}, \bibinfo {author}
  {\bibfnamefont {D.}~\bibnamefont {Hunger}}, \ and\ \bibinfo {author}
  {\bibfnamefont {J.}~\bibnamefont {Reichel}},\ }\href@noop {} {\bibfield
  {journal} {\bibinfo  {journal} {Nature}\ }\textbf {\bibinfo {volume} {450}},\
  \bibinfo {pages} {272} (\bibinfo {year} {2007})}\BibitemShut {NoStop}%
\bibitem [{\citenamefont {Gupta}\ \emph {et~al.}(2007)\citenamefont {Gupta},
  \citenamefont {Moore}, \citenamefont {Murch},\ and\ \citenamefont
  {Stamper-Kurn}}]{st_kurn_2007}%
  \BibitemOpen
  \bibfield  {author} {\bibinfo {author} {\bibfnamefont {S.}~\bibnamefont
  {Gupta}}, \bibinfo {author} {\bibfnamefont {K.~L.}\ \bibnamefont {Moore}},
  \bibinfo {author} {\bibfnamefont {K.~W.}\ \bibnamefont {Murch}}, \ and\
  \bibinfo {author} {\bibfnamefont {D.~M.}\ \bibnamefont {Stamper-Kurn}},\
  }\href {\doibase 10.1103/PhysRevLett.99.213601} {\bibfield  {journal}
  {\bibinfo  {journal} {Phys. Rev. Lett.}\ }\textbf {\bibinfo {volume} {99}},\
  \bibinfo {pages} {213601} (\bibinfo {year} {2007})}\BibitemShut {NoStop}%
\bibitem [{\citenamefont {Baumann}\ \emph {et~al.}(2010)\citenamefont
  {Baumann}, \citenamefont {Guerlin}, \citenamefont {Brennecke},\ and\
  \citenamefont {Esslinger}}]{eth_2010}%
  \BibitemOpen
  \bibfield  {author} {\bibinfo {author} {\bibfnamefont {K.}~\bibnamefont
  {Baumann}}, \bibinfo {author} {\bibfnamefont {C.}~\bibnamefont {Guerlin}},
  \bibinfo {author} {\bibfnamefont {F.}~\bibnamefont {Brennecke}}, \ and\
  \bibinfo {author} {\bibfnamefont {T.}~\bibnamefont {Esslinger}},\ }\href@noop
  {} {\bibfield  {journal} {\bibinfo  {journal} {Nature}\ }\textbf {\bibinfo
  {volume} {464}},\ \bibinfo {pages} {1301} (\bibinfo {year}
  {2010})}\BibitemShut {NoStop}%
\bibitem [{\citenamefont {Arnold}\ \emph {et~al.}(2012)\citenamefont {Arnold},
  \citenamefont {Baden},\ and\ \citenamefont {Barrett}}]{barrett_2012}%
  \BibitemOpen
  \bibfield  {author} {\bibinfo {author} {\bibfnamefont {K.~J.}\ \bibnamefont
  {Arnold}}, \bibinfo {author} {\bibfnamefont {M.~P.}\ \bibnamefont {Baden}}, \
  and\ \bibinfo {author} {\bibfnamefont {M.~D.}\ \bibnamefont {Barrett}},\
  }\href {\doibase 10.1103/PhysRevLett.109.153002} {\bibfield  {journal}
  {\bibinfo  {journal} {Phys. Rev. Lett.}\ }\textbf {\bibinfo {volume} {109}},\
  \bibinfo {pages} {153002} (\bibinfo {year} {2012})}\BibitemShut {NoStop}%
\bibitem [{\citenamefont {Ke\ss{}ler}\ \emph {et~al.}(2014)\citenamefont
  {Ke\ss{}ler}, \citenamefont {Klinder}, \citenamefont {Wolke},\ and\
  \citenamefont {Hemmerich}}]{hemmerich_SR_2014}%
  \BibitemOpen
  \bibfield  {author} {\bibinfo {author} {\bibfnamefont {H.}~\bibnamefont
  {Ke\ss{}ler}}, \bibinfo {author} {\bibfnamefont {J.}~\bibnamefont {Klinder}},
  \bibinfo {author} {\bibfnamefont {M.}~\bibnamefont {Wolke}}, \ and\ \bibinfo
  {author} {\bibfnamefont {A.}~\bibnamefont {Hemmerich}},\ }\href {\doibase
  10.1103/PhysRevLett.113.070404} {\bibfield  {journal} {\bibinfo  {journal}
  {Phys. Rev. Lett.}\ }\textbf {\bibinfo {volume} {113}},\ \bibinfo {pages}
  {070404} (\bibinfo {year} {2014})}\BibitemShut {NoStop}%
\bibitem [{\citenamefont {Kollar}\ \emph {et~al.}(2015)\citenamefont {Kollar},
  \citenamefont {Papageorge}, \citenamefont {Baumann}, \citenamefont {Armen},\
  and\ \citenamefont {Lev}}]{lev_multimode_exp_2015}%
  \BibitemOpen
  \bibfield  {author} {\bibinfo {author} {\bibfnamefont {A.~J.}\ \bibnamefont
  {Kollar}}, \bibinfo {author} {\bibfnamefont {A.~T.}\ \bibnamefont
  {Papageorge}}, \bibinfo {author} {\bibfnamefont {K.}~\bibnamefont {Baumann}},
  \bibinfo {author} {\bibfnamefont {M.~A.}\ \bibnamefont {Armen}}, \ and\
  \bibinfo {author} {\bibfnamefont {B.~L.}\ \bibnamefont {Lev}},\ }\href
  {http://stacks.iop.org/1367-2630/17/i=4/a=043012} {\bibfield  {journal}
  {\bibinfo  {journal} {New Journal of Physics}\ }\textbf {\bibinfo {volume}
  {17}},\ \bibinfo {pages} {043012} (\bibinfo {year} {2015})}\BibitemShut
  {NoStop}%
\bibitem [{\citenamefont {Ritsch}\ \emph {et~al.}(2013)\citenamefont {Ritsch},
  \citenamefont {Domokos}, \citenamefont {Brennecke},\ and\ \citenamefont
  {Esslinger}}]{cavity_rmp}%
  \BibitemOpen
  \bibfield  {author} {\bibinfo {author} {\bibfnamefont {H.}~\bibnamefont
  {Ritsch}}, \bibinfo {author} {\bibfnamefont {P.}~\bibnamefont {Domokos}},
  \bibinfo {author} {\bibfnamefont {F.}~\bibnamefont {Brennecke}}, \ and\
  \bibinfo {author} {\bibfnamefont {T.}~\bibnamefont {Esslinger}},\ }\href
  {\doibase 10.1103/RevModPhys.85.553} {\bibfield  {journal} {\bibinfo
  {journal} {Rev. Mod. Phys.}\ }\textbf {\bibinfo {volume} {85}},\ \bibinfo
  {pages} {553} (\bibinfo {year} {2013})}\BibitemShut {NoStop}%
\bibitem [{\citenamefont {Maschler}\ \emph {et~al.}(2008)\citenamefont
  {Maschler}, \citenamefont {Mekhov},\ and\ \citenamefont
  {Ritsch}}]{maschler_2008}%
  \BibitemOpen
  \bibfield  {author} {\bibinfo {author} {\bibfnamefont {C.}~\bibnamefont
  {Maschler}}, \bibinfo {author} {\bibfnamefont {I.~B.}\ \bibnamefont
  {Mekhov}}, \ and\ \bibinfo {author} {\bibfnamefont {H.}~\bibnamefont
  {Ritsch}},\ }\href {\doibase 10.1140/epjd/e2008-00016-4} {\bibfield
  {journal} {\bibinfo  {journal} {The European Physical Journal D}\ }\textbf
  {\bibinfo {volume} {46}},\ \bibinfo {pages} {545} (\bibinfo {year}
  {2008})}\BibitemShut {NoStop}%
\bibitem [{\citenamefont {Gopalakrishnan}\ \emph {et~al.}(2009)\citenamefont
  {Gopalakrishnan}, \citenamefont {Lev},\ and\ \citenamefont
  {Goldbart}}]{gopal_2009}%
  \BibitemOpen
  \bibfield  {author} {\bibinfo {author} {\bibfnamefont {S.}~\bibnamefont
  {Gopalakrishnan}}, \bibinfo {author} {\bibfnamefont {B.~L.}\ \bibnamefont
  {Lev}}, \ and\ \bibinfo {author} {\bibfnamefont {P.~M.}\ \bibnamefont
  {Goldbart}},\ }\href@noop {} {\bibfield  {journal} {\bibinfo  {journal}
  {Nature Physics}\ }\textbf {\bibinfo {volume} {5}},\ \bibinfo {pages} {845}
  (\bibinfo {year} {2009})}\BibitemShut {NoStop}%
\bibitem [{\citenamefont {Gopalakrishnan}\ \emph {et~al.}(2011)\citenamefont
  {Gopalakrishnan}, \citenamefont {Lev},\ and\ \citenamefont
  {Goldbart}}]{gopal_glass_2011}%
  \BibitemOpen
  \bibfield  {author} {\bibinfo {author} {\bibfnamefont {S.}~\bibnamefont
  {Gopalakrishnan}}, \bibinfo {author} {\bibfnamefont {B.~L.}\ \bibnamefont
  {Lev}}, \ and\ \bibinfo {author} {\bibfnamefont {P.~M.}\ \bibnamefont
  {Goldbart}},\ }\href {\doibase 10.1103/PhysRevLett.107.277201} {\bibfield
  {journal} {\bibinfo  {journal} {Phys. Rev. Lett.}\ }\textbf {\bibinfo
  {volume} {107}},\ \bibinfo {pages} {277201} (\bibinfo {year}
  {2011})}\BibitemShut {NoStop}%
\bibitem [{\citenamefont {Strack}\ and\ \citenamefont
  {Sachdev}(2011)}]{strack_2011}%
  \BibitemOpen
  \bibfield  {author} {\bibinfo {author} {\bibfnamefont {P.}~\bibnamefont
  {Strack}}\ and\ \bibinfo {author} {\bibfnamefont {S.}~\bibnamefont
  {Sachdev}},\ }\href {\doibase 10.1103/PhysRevLett.107.277202} {\bibfield
  {journal} {\bibinfo  {journal} {Phys. Rev. Lett.}\ }\textbf {\bibinfo
  {volume} {107}},\ \bibinfo {pages} {277202} (\bibinfo {year}
  {2011})}\BibitemShut {NoStop}%
\bibitem [{\citenamefont {Caballero-Benitez}\ and\ \citenamefont
  {Mekhov}(2015)}]{caballero_2015}%
  \BibitemOpen
  \bibfield  {author} {\bibinfo {author} {\bibfnamefont {S.~F.}\ \bibnamefont
  {Caballero-Benitez}}\ and\ \bibinfo {author} {\bibfnamefont {I.~B.}\
  \bibnamefont {Mekhov}},\ }\href {\doibase 10.1103/PhysRevLett.115.243604}
  {\bibfield  {journal} {\bibinfo  {journal} {Phys. Rev. Lett.}\ }\textbf
  {\bibinfo {volume} {115}},\ \bibinfo {pages} {243604} (\bibinfo {year}
  {2015})}\BibitemShut {NoStop}%
\bibitem [{\citenamefont {Mazzucchi}\ \emph {et~al.}(2016)\citenamefont
  {Mazzucchi}, \citenamefont {Kozlowski}, \citenamefont {Caballero-Benitez},
  \citenamefont {Elliott},\ and\ \citenamefont
  {Mekhov}}]{mazzucchi2016quantum}%
  \BibitemOpen
  \bibfield  {author} {\bibinfo {author} {\bibfnamefont {G.}~\bibnamefont
  {Mazzucchi}}, \bibinfo {author} {\bibfnamefont {W.}~\bibnamefont
  {Kozlowski}}, \bibinfo {author} {\bibfnamefont {S.~F.}\ \bibnamefont
  {Caballero-Benitez}}, \bibinfo {author} {\bibfnamefont {T.~J.}\ \bibnamefont
  {Elliott}}, \ and\ \bibinfo {author} {\bibfnamefont {I.~B.}\ \bibnamefont
  {Mekhov}},\ }\href@noop {} {\bibfield  {journal} {\bibinfo  {journal}
  {Physical Review A}\ }\textbf {\bibinfo {volume} {93}},\ \bibinfo {pages}
  {023632} (\bibinfo {year} {2016})}\BibitemShut {NoStop}%
\bibitem [{\citenamefont {Torggler}\ \emph {et~al.}(2016)\citenamefont
  {Torggler}, \citenamefont {Kr{\"a}mer},\ and\ \citenamefont
  {Ritsch}}]{torggler2016quantum}%
  \BibitemOpen
  \bibfield  {author} {\bibinfo {author} {\bibfnamefont {V.}~\bibnamefont
  {Torggler}}, \bibinfo {author} {\bibfnamefont {S.}~\bibnamefont
  {Kr{\"a}mer}}, \ and\ \bibinfo {author} {\bibfnamefont {H.}~\bibnamefont
  {Ritsch}},\ }\href@noop {} {\bibfield  {journal} {\bibinfo  {journal} {arXiv
  preprint arXiv:1609.06250}\ } (\bibinfo {year} {2016})}\BibitemShut {NoStop}%
\bibitem [{\citenamefont {{Koll{\'a}r}}\ \emph {et~al.}(2016)\citenamefont
  {{Koll{\'a}r}}, \citenamefont {{Papageorge}}, \citenamefont {{Vaidya}},
  \citenamefont {{Guo}}, \citenamefont {{Keeling}},\ and\ \citenamefont
  {{Lev}}}]{lev_supermode}%
  \BibitemOpen
  \bibfield  {author} {\bibinfo {author} {\bibfnamefont {A.~J.}\ \bibnamefont
  {{Koll{\'a}r}}}, \bibinfo {author} {\bibfnamefont {A.~T.}\ \bibnamefont
  {{Papageorge}}}, \bibinfo {author} {\bibfnamefont {V.~D.}\ \bibnamefont
  {{Vaidya}}}, \bibinfo {author} {\bibfnamefont {Y.}~\bibnamefont {{Guo}}},
  \bibinfo {author} {\bibfnamefont {J.}~\bibnamefont {{Keeling}}}, \ and\
  \bibinfo {author} {\bibfnamefont {B.~L.}\ \bibnamefont {{Lev}}},\ }\href@noop
  {} {\bibfield  {journal} {\bibinfo  {journal} {ArXiv e-prints}\ } (\bibinfo
  {year} {2016})},\ \Eprint {http://arxiv.org/abs/1606.04127} {arXiv:1606.04127
  [cond-mat.quant-gas]} \BibitemShut {NoStop}%
\bibitem [{\citenamefont {Domokos}\ and\ \citenamefont
  {Ritsch}(2002)}]{ritsch_2002}%
  \BibitemOpen
  \bibfield  {author} {\bibinfo {author} {\bibfnamefont {P.}~\bibnamefont
  {Domokos}}\ and\ \bibinfo {author} {\bibfnamefont {H.}~\bibnamefont
  {Ritsch}},\ }\href {\doibase 10.1103/PhysRevLett.89.253003} {\bibfield
  {journal} {\bibinfo  {journal} {Phys. Rev. Lett.}\ }\textbf {\bibinfo
  {volume} {89}},\ \bibinfo {pages} {253003} (\bibinfo {year}
  {2002})}\BibitemShut {NoStop}%
\bibitem [{\citenamefont {Nagy}\ \emph {et~al.}(2008)\citenamefont {Nagy},
  \citenamefont {Szirmai},\ and\ \citenamefont {Domokos}}]{domokos_2008}%
  \BibitemOpen
  \bibfield  {author} {\bibinfo {author} {\bibfnamefont {D.}~\bibnamefont
  {Nagy}}, \bibinfo {author} {\bibfnamefont {G.}~\bibnamefont {Szirmai}}, \
  and\ \bibinfo {author} {\bibfnamefont {P.}~\bibnamefont {Domokos}},\ }\href
  {\doibase 10.1140/epjd/e2008-00074-6} {\bibfield  {journal} {\bibinfo
  {journal} {The European Physical Journal D}\ }\textbf {\bibinfo {volume}
  {48}},\ \bibinfo {pages} {127} (\bibinfo {year} {2008})}\BibitemShut
  {NoStop}%
\bibitem [{\citenamefont {Piazza}\ \emph {et~al.}(2013)\citenamefont {Piazza},
  \citenamefont {Strack},\ and\ \citenamefont {Zwerger}}]{piazza_bose}%
  \BibitemOpen
  \bibfield  {author} {\bibinfo {author} {\bibfnamefont {F.}~\bibnamefont
  {Piazza}}, \bibinfo {author} {\bibfnamefont {P.}~\bibnamefont {Strack}}, \
  and\ \bibinfo {author} {\bibfnamefont {W.}~\bibnamefont {Zwerger}},\ }\href
  {\doibase http://dx.doi.org/10.1016/j.aop.2013.08.015} {\bibfield  {journal}
  {\bibinfo  {journal} {Annals of Physics}\ }\textbf {\bibinfo {volume}
  {339}},\ \bibinfo {pages} {135 } (\bibinfo {year} {2013})}\BibitemShut
  {NoStop}%
\bibitem [{\citenamefont {Lode}\ and\ \citenamefont
  {Bruder}(2016)}]{lode2016fragmented}%
  \BibitemOpen
  \bibfield  {author} {\bibinfo {author} {\bibfnamefont {A.~U.}\ \bibnamefont
  {Lode}}\ and\ \bibinfo {author} {\bibfnamefont {C.}~\bibnamefont {Bruder}},\
  }\href@noop {} {\bibfield  {journal} {\bibinfo  {journal} {arXiv preprint
  arXiv:1606.06058}\ } (\bibinfo {year} {2016})}\BibitemShut {NoStop}%
\bibitem [{\citenamefont {Keeling}\ \emph {et~al.}(2014)\citenamefont
  {Keeling}, \citenamefont {Bhaseen},\ and\ \citenamefont
  {Simons}}]{keeling_2014}%
  \BibitemOpen
  \bibfield  {author} {\bibinfo {author} {\bibfnamefont {J.}~\bibnamefont
  {Keeling}}, \bibinfo {author} {\bibfnamefont {J.}~\bibnamefont {Bhaseen}}, \
  and\ \bibinfo {author} {\bibfnamefont {B.}~\bibnamefont {Simons}},\ }\href
  {\doibase 10.1103/PhysRevLett.112.143002} {\bibfield  {journal} {\bibinfo
  {journal} {Phys. Rev. Lett.}\ }\textbf {\bibinfo {volume} {112}},\ \bibinfo
  {pages} {143002} (\bibinfo {year} {2014})}\BibitemShut {NoStop}%
\bibitem [{\citenamefont {Chen}\ \emph {et~al.}(2014)\citenamefont {Chen},
  \citenamefont {Yu},\ and\ \citenamefont {Zhai}}]{zhai_2014}%
  \BibitemOpen
  \bibfield  {author} {\bibinfo {author} {\bibfnamefont {Y.}~\bibnamefont
  {Chen}}, \bibinfo {author} {\bibfnamefont {Z.}~\bibnamefont {Yu}}, \ and\
  \bibinfo {author} {\bibfnamefont {H.}~\bibnamefont {Zhai}},\ }\href {\doibase
  10.1103/PhysRevLett.112.143004} {\bibfield  {journal} {\bibinfo  {journal}
  {Phys. Rev. Lett.}\ }\textbf {\bibinfo {volume} {112}},\ \bibinfo {pages}
  {143004} (\bibinfo {year} {2014})}\BibitemShut {NoStop}%
\bibitem [{\citenamefont {Piazza}\ and\ \citenamefont
  {Strack}(2014{\natexlab{a}})}]{piazza_fermi}%
  \BibitemOpen
  \bibfield  {author} {\bibinfo {author} {\bibfnamefont {F.}~\bibnamefont
  {Piazza}}\ and\ \bibinfo {author} {\bibfnamefont {P.}~\bibnamefont
  {Strack}},\ }\href {\doibase 10.1103/PhysRevLett.112.143003} {\bibfield
  {journal} {\bibinfo  {journal} {Phys. Rev. Lett.}\ }\textbf {\bibinfo
  {volume} {112}},\ \bibinfo {pages} {143003} (\bibinfo {year}
  {2014}{\natexlab{a}})}\BibitemShut {NoStop}%
\bibitem [{\citenamefont {Piazza}\ and\ \citenamefont
  {Strack}(2014{\natexlab{b}})}]{piazza_QKE}%
  \BibitemOpen
  \bibfield  {author} {\bibinfo {author} {\bibfnamefont {F.}~\bibnamefont
  {Piazza}}\ and\ \bibinfo {author} {\bibfnamefont {P.}~\bibnamefont
  {Strack}},\ }\href {\doibase 10.1103/PhysRevA.90.043823} {\bibfield
  {journal} {\bibinfo  {journal} {Phys. Rev. A}\ }\textbf {\bibinfo {volume}
  {90}},\ \bibinfo {pages} {043823} (\bibinfo {year}
  {2014}{\natexlab{b}})}\BibitemShut {NoStop}%
\bibitem [{\citenamefont {Sandner}\ \emph {et~al.}(2015)\citenamefont
  {Sandner}, \citenamefont {Niedenzu}, \citenamefont {Piazza},\ and\
  \citenamefont {Ritsch}}]{sandner_fermi}%
  \BibitemOpen
  \bibfield  {author} {\bibinfo {author} {\bibfnamefont {R.~M.}\ \bibnamefont
  {Sandner}}, \bibinfo {author} {\bibfnamefont {W.}~\bibnamefont {Niedenzu}},
  \bibinfo {author} {\bibfnamefont {F.}~\bibnamefont {Piazza}}, \ and\ \bibinfo
  {author} {\bibfnamefont {H.}~\bibnamefont {Ritsch}},\ }\href
  {http://stacks.iop.org/0295-5075/111/i=5/a=53001} {\bibfield  {journal}
  {\bibinfo  {journal} {EPL (Europhysics Letters)}\ }\textbf {\bibinfo {volume}
  {111}},\ \bibinfo {pages} {53001} (\bibinfo {year} {2015})}\BibitemShut
  {NoStop}%
\bibitem [{\citenamefont {Habibian}\ \emph {et~al.}(2013)\citenamefont
  {Habibian}, \citenamefont {Winter}, \citenamefont {Paganelli}, \citenamefont
  {Rieger},\ and\ \citenamefont {Morigi}}]{morigi_2013}%
  \BibitemOpen
  \bibfield  {author} {\bibinfo {author} {\bibfnamefont {H.}~\bibnamefont
  {Habibian}}, \bibinfo {author} {\bibfnamefont {A.}~\bibnamefont {Winter}},
  \bibinfo {author} {\bibfnamefont {S.}~\bibnamefont {Paganelli}}, \bibinfo
  {author} {\bibfnamefont {H.}~\bibnamefont {Rieger}}, \ and\ \bibinfo {author}
  {\bibfnamefont {G.}~\bibnamefont {Morigi}},\ }\href {\doibase
  10.1103/PhysRevLett.110.075304} {\bibfield  {journal} {\bibinfo  {journal}
  {Phys. Rev. Lett.}\ }\textbf {\bibinfo {volume} {110}},\ \bibinfo {pages}
  {075304} (\bibinfo {year} {2013})}\BibitemShut {NoStop}%
\bibitem [{\citenamefont {Li}\ \emph {et~al.}(2013)\citenamefont {Li},
  \citenamefont {He},\ and\ \citenamefont {Hofstetter}}]{hofstetter13}%
  \BibitemOpen
  \bibfield  {author} {\bibinfo {author} {\bibfnamefont {Y.}~\bibnamefont
  {Li}}, \bibinfo {author} {\bibfnamefont {L.}~\bibnamefont {He}}, \ and\
  \bibinfo {author} {\bibfnamefont {W.}~\bibnamefont {Hofstetter}},\ }\href
  {\doibase 10.1103/PhysRevA.87.051604} {\bibfield  {journal} {\bibinfo
  {journal} {Phys. Rev. A}\ }\textbf {\bibinfo {volume} {87}},\ \bibinfo
  {pages} {051604} (\bibinfo {year} {2013})}\BibitemShut {NoStop}%
\bibitem [{\citenamefont {Bakhtiari}\ \emph {et~al.}(2015)\citenamefont
  {Bakhtiari}, \citenamefont {Hemmerich}, \citenamefont {Ritsch},\ and\
  \citenamefont {Thorwart}}]{bakhtiari_2015}%
  \BibitemOpen
  \bibfield  {author} {\bibinfo {author} {\bibfnamefont {M.~R.}\ \bibnamefont
  {Bakhtiari}}, \bibinfo {author} {\bibfnamefont {A.}~\bibnamefont
  {Hemmerich}}, \bibinfo {author} {\bibfnamefont {H.}~\bibnamefont {Ritsch}}, \
  and\ \bibinfo {author} {\bibfnamefont {M.}~\bibnamefont {Thorwart}},\ }\href
  {\doibase 10.1103/PhysRevLett.114.123601} {\bibfield  {journal} {\bibinfo
  {journal} {Phys. Rev. Lett.}\ }\textbf {\bibinfo {volume} {114}},\ \bibinfo
  {pages} {123601} (\bibinfo {year} {2015})}\BibitemShut {NoStop}%
\bibitem [{\citenamefont {Chen}\ \emph {et~al.}(2016)\citenamefont {Chen},
  \citenamefont {Yu},\ and\ \citenamefont {Zhai}}]{zhai_mottcav_2016}%
  \BibitemOpen
  \bibfield  {author} {\bibinfo {author} {\bibfnamefont {Y.}~\bibnamefont
  {Chen}}, \bibinfo {author} {\bibfnamefont {Z.}~\bibnamefont {Yu}}, \ and\
  \bibinfo {author} {\bibfnamefont {H.}~\bibnamefont {Zhai}},\ }\href {\doibase
  10.1103/PhysRevA.93.041601} {\bibfield  {journal} {\bibinfo  {journal} {Phys.
  Rev. A}\ }\textbf {\bibinfo {volume} {93}},\ \bibinfo {pages} {041601}
  (\bibinfo {year} {2016})}\BibitemShut {NoStop}%
\bibitem [{\citenamefont {Dogra}\ \emph {et~al.}(2016)\citenamefont {Dogra},
  \citenamefont {Brennecke}, \citenamefont {Huber},\ and\ \citenamefont
  {Donner}}]{dogra2016phase}%
  \BibitemOpen
  \bibfield  {author} {\bibinfo {author} {\bibfnamefont {N.}~\bibnamefont
  {Dogra}}, \bibinfo {author} {\bibfnamefont {F.}~\bibnamefont {Brennecke}},
  \bibinfo {author} {\bibfnamefont {S.~D.}\ \bibnamefont {Huber}}, \ and\
  \bibinfo {author} {\bibfnamefont {T.}~\bibnamefont {Donner}},\ }\href
  {\doibase 10.1103/PhysRevA.94.023632} {\bibfield  {journal} {\bibinfo
  {journal} {Phys. Rev. A}\ }\textbf {\bibinfo {volume} {94}},\ \bibinfo
  {pages} {023632} (\bibinfo {year} {2016})}\BibitemShut {NoStop}%
\bibitem [{\citenamefont {Niederle}\ \emph {et~al.}(2016)\citenamefont
  {Niederle}, \citenamefont {Morigi},\ and\ \citenamefont
  {Rieger}}]{morigi_competing_2016}%
  \BibitemOpen
  \bibfield  {author} {\bibinfo {author} {\bibfnamefont {A.~E.}\ \bibnamefont
  {Niederle}}, \bibinfo {author} {\bibfnamefont {G.}~\bibnamefont {Morigi}}, \
  and\ \bibinfo {author} {\bibfnamefont {H.}~\bibnamefont {Rieger}},\ }\href
  {\doibase 10.1103/PhysRevA.94.033607} {\bibfield  {journal} {\bibinfo
  {journal} {Phys. Rev. A}\ }\textbf {\bibinfo {volume} {94}},\ \bibinfo
  {pages} {033607} (\bibinfo {year} {2016})}\BibitemShut {NoStop}%
\bibitem [{\citenamefont {Gelhausen}\ \emph {et~al.}(2016)\citenamefont
  {Gelhausen}, \citenamefont {Buchhold}, \citenamefont {Rosch},\ and\
  \citenamefont {Strack}}]{gelhausen2016quantum}%
  \BibitemOpen
  \bibfield  {author} {\bibinfo {author} {\bibfnamefont {J.}~\bibnamefont
  {Gelhausen}}, \bibinfo {author} {\bibfnamefont {M.}~\bibnamefont {Buchhold}},
  \bibinfo {author} {\bibfnamefont {A.}~\bibnamefont {Rosch}}, \ and\ \bibinfo
  {author} {\bibfnamefont {P.}~\bibnamefont {Strack}},\ }\href {\doibase
  10.21468/SciPostPhys.1.1.004} {\bibfield  {journal} {\bibinfo  {journal}
  {SciPost Phys.}\ }\textbf {\bibinfo {volume} {1}},\ \bibinfo {pages} {004}
  (\bibinfo {year} {2016})}\BibitemShut {NoStop}%
\bibitem [{\citenamefont {Klinder}\ \emph {et~al.}(2015)\citenamefont
  {Klinder}, \citenamefont {Ke\ss{}ler}, \citenamefont {Bakhtiari},
  \citenamefont {Thorwart},\ and\ \citenamefont
  {Hemmerich}}]{hemmerich_mottcav_2015}%
  \BibitemOpen
  \bibfield  {author} {\bibinfo {author} {\bibfnamefont {J.}~\bibnamefont
  {Klinder}}, \bibinfo {author} {\bibfnamefont {H.}~\bibnamefont {Ke\ss{}ler}},
  \bibinfo {author} {\bibfnamefont {M.~R.}\ \bibnamefont {Bakhtiari}}, \bibinfo
  {author} {\bibfnamefont {M.}~\bibnamefont {Thorwart}}, \ and\ \bibinfo
  {author} {\bibfnamefont {A.}~\bibnamefont {Hemmerich}},\ }\href {\doibase
  10.1103/PhysRevLett.115.230403} {\bibfield  {journal} {\bibinfo  {journal}
  {Phys. Rev. Lett.}\ }\textbf {\bibinfo {volume} {115}},\ \bibinfo {pages}
  {230403} (\bibinfo {year} {2015})}\BibitemShut {NoStop}%
\bibitem [{\citenamefont {Landig}\ \emph {et~al.}(2016)\citenamefont {Landig},
  \citenamefont {Hruby}, \citenamefont {Dogra}, \citenamefont {Landini},
  \citenamefont {Mottl}, \citenamefont {Donner},\ and\ \citenamefont
  {Esslinger}}]{landig2016quantum}%
  \BibitemOpen
  \bibfield  {author} {\bibinfo {author} {\bibfnamefont {R.}~\bibnamefont
  {Landig}}, \bibinfo {author} {\bibfnamefont {L.}~\bibnamefont {Hruby}},
  \bibinfo {author} {\bibfnamefont {N.}~\bibnamefont {Dogra}}, \bibinfo
  {author} {\bibfnamefont {M.}~\bibnamefont {Landini}}, \bibinfo {author}
  {\bibfnamefont {R.}~\bibnamefont {Mottl}}, \bibinfo {author} {\bibfnamefont
  {T.}~\bibnamefont {Donner}}, \ and\ \bibinfo {author} {\bibfnamefont
  {T.}~\bibnamefont {Esslinger}},\ }\href
  {http://dx.doi.org/10.1038/nature17409} {\bibfield  {journal} {\bibinfo
  {journal} {Nature}\ }\textbf {\bibinfo {volume} {532}},\ \bibinfo {pages}
  {476} (\bibinfo {year} {2016})}\BibitemShut {NoStop}%
\bibitem [{\citenamefont {L{\'e}onard}\ \emph {et~al.}(2016)\citenamefont
  {L{\'e}onard}, \citenamefont {Morales}, \citenamefont {Zupancic},
  \citenamefont {Esslinger},\ and\ \citenamefont
  {Donner}}]{leonard2016supersolid}%
  \BibitemOpen
  \bibfield  {author} {\bibinfo {author} {\bibfnamefont {J.}~\bibnamefont
  {L{\'e}onard}}, \bibinfo {author} {\bibfnamefont {A.}~\bibnamefont
  {Morales}}, \bibinfo {author} {\bibfnamefont {P.}~\bibnamefont {Zupancic}},
  \bibinfo {author} {\bibfnamefont {T.}~\bibnamefont {Esslinger}}, \ and\
  \bibinfo {author} {\bibfnamefont {T.}~\bibnamefont {Donner}},\ }\href@noop {}
  {\bibfield  {journal} {\bibinfo  {journal} {arXiv preprint arXiv:1609.09053}\
  } (\bibinfo {year} {2016})}\BibitemShut {NoStop}%
\bibitem [{\citenamefont {Mivehvar}\ and\ \citenamefont
  {Feder}(2014)}]{mivehvar2014synthetic}%
  \BibitemOpen
  \bibfield  {author} {\bibinfo {author} {\bibfnamefont {F.}~\bibnamefont
  {Mivehvar}}\ and\ \bibinfo {author} {\bibfnamefont {D.~L.}\ \bibnamefont
  {Feder}},\ }\href@noop {} {\bibfield  {journal} {\bibinfo  {journal}
  {Physical Review A}\ }\textbf {\bibinfo {volume} {89}},\ \bibinfo {pages}
  {013803} (\bibinfo {year} {2014})}\BibitemShut {NoStop}%
\bibitem [{\citenamefont {Deng}\ \emph {et~al.}(2014)\citenamefont {Deng},
  \citenamefont {Cheng}, \citenamefont {Jing},\ and\ \citenamefont
  {Yi}}]{sr_soc_yi_2014}%
  \BibitemOpen
  \bibfield  {author} {\bibinfo {author} {\bibfnamefont {Y.}~\bibnamefont
  {Deng}}, \bibinfo {author} {\bibfnamefont {J.}~\bibnamefont {Cheng}},
  \bibinfo {author} {\bibfnamefont {H.}~\bibnamefont {Jing}}, \ and\ \bibinfo
  {author} {\bibfnamefont {S.}~\bibnamefont {Yi}},\ }\href {\doibase
  10.1103/PhysRevLett.112.143007} {\bibfield  {journal} {\bibinfo  {journal}
  {Phys. Rev. Lett.}\ }\textbf {\bibinfo {volume} {112}},\ \bibinfo {pages}
  {143007} (\bibinfo {year} {2014})}\BibitemShut {NoStop}%
\bibitem [{\citenamefont {Padhi}\ and\ \citenamefont
  {Ghosh}(2014)}]{cavity_soc_Gosh_2014}%
  \BibitemOpen
  \bibfield  {author} {\bibinfo {author} {\bibfnamefont {B.}~\bibnamefont
  {Padhi}}\ and\ \bibinfo {author} {\bibfnamefont {S.}~\bibnamefont {Ghosh}},\
  }\href {\doibase 10.1103/PhysRevA.90.023627} {\bibfield  {journal} {\bibinfo
  {journal} {Phys. Rev. A}\ }\textbf {\bibinfo {volume} {90}},\ \bibinfo
  {pages} {023627} (\bibinfo {year} {2014})}\BibitemShut {NoStop}%
\bibitem [{\citenamefont {Dong}\ \emph {et~al.}(2014)\citenamefont {Dong},
  \citenamefont {Zhou}, \citenamefont {Wu}, \citenamefont {Ramachandhran},\
  and\ \citenamefont {Pu}}]{cavity_soc_Han_2014}%
  \BibitemOpen
  \bibfield  {author} {\bibinfo {author} {\bibfnamefont {L.}~\bibnamefont
  {Dong}}, \bibinfo {author} {\bibfnamefont {L.}~\bibnamefont {Zhou}}, \bibinfo
  {author} {\bibfnamefont {B.}~\bibnamefont {Wu}}, \bibinfo {author}
  {\bibfnamefont {B.}~\bibnamefont {Ramachandhran}}, \ and\ \bibinfo {author}
  {\bibfnamefont {H.}~\bibnamefont {Pu}},\ }\href {\doibase
  10.1103/PhysRevA.89.011602} {\bibfield  {journal} {\bibinfo  {journal} {Phys.
  Rev. A}\ }\textbf {\bibinfo {volume} {89}},\ \bibinfo {pages} {011602}
  (\bibinfo {year} {2014})}\BibitemShut {NoStop}%
\bibitem [{\citenamefont {Pan}\ \emph {et~al.}(2015)\citenamefont {Pan},
  \citenamefont {Liu}, \citenamefont {Zhang}, \citenamefont {Yi},\ and\
  \citenamefont {Guo}}]{sr_soc_fermi_guo_2015}%
  \BibitemOpen
  \bibfield  {author} {\bibinfo {author} {\bibfnamefont {J.-S.}\ \bibnamefont
  {Pan}}, \bibinfo {author} {\bibfnamefont {X.-J.}\ \bibnamefont {Liu}},
  \bibinfo {author} {\bibfnamefont {W.}~\bibnamefont {Zhang}}, \bibinfo
  {author} {\bibfnamefont {W.}~\bibnamefont {Yi}}, \ and\ \bibinfo {author}
  {\bibfnamefont {G.-C.}\ \bibnamefont {Guo}},\ }\href {\doibase
  10.1103/PhysRevLett.115.045303} {\bibfield  {journal} {\bibinfo  {journal}
  {Phys. Rev. Lett.}\ }\textbf {\bibinfo {volume} {115}},\ \bibinfo {pages}
  {045303} (\bibinfo {year} {2015})}\BibitemShut {NoStop}%
\bibitem [{\citenamefont {Kollath}\ \emph {et~al.}(2016)\citenamefont
  {Kollath}, \citenamefont {Sheikhan}, \citenamefont {Wolff},\ and\
  \citenamefont {Brennecke}}]{kollath2016ultracold}%
  \BibitemOpen
  \bibfield  {author} {\bibinfo {author} {\bibfnamefont {C.}~\bibnamefont
  {Kollath}}, \bibinfo {author} {\bibfnamefont {A.}~\bibnamefont {Sheikhan}},
  \bibinfo {author} {\bibfnamefont {S.}~\bibnamefont {Wolff}}, \ and\ \bibinfo
  {author} {\bibfnamefont {F.}~\bibnamefont {Brennecke}},\ }\href@noop {}
  {\bibfield  {journal} {\bibinfo  {journal} {Physical review letters}\
  }\textbf {\bibinfo {volume} {116}},\ \bibinfo {pages} {060401} (\bibinfo
  {year} {2016})}\BibitemShut {NoStop}%
\bibitem [{\citenamefont {Sheikhan}\ \emph {et~al.}(2016)\citenamefont
  {Sheikhan}, \citenamefont {Brennecke},\ and\ \citenamefont
  {Kollath}}]{sheikhan2016cavity}%
  \BibitemOpen
  \bibfield  {author} {\bibinfo {author} {\bibfnamefont {A.}~\bibnamefont
  {Sheikhan}}, \bibinfo {author} {\bibfnamefont {F.}~\bibnamefont {Brennecke}},
  \ and\ \bibinfo {author} {\bibfnamefont {C.}~\bibnamefont {Kollath}},\
  }\href@noop {} {\bibfield  {journal} {\bibinfo  {journal} {Physical Review
  A}\ }\textbf {\bibinfo {volume} {93}},\ \bibinfo {pages} {043609} (\bibinfo
  {year} {2016})}\BibitemShut {NoStop}%
\bibitem [{\citenamefont {Gul\'acsi}\ and\ \citenamefont
  {D\'ora}(2015)}]{dicke_spinHall_2015}%
  \BibitemOpen
  \bibfield  {author} {\bibinfo {author} {\bibfnamefont {B.}~\bibnamefont
  {Gul\'acsi}}\ and\ \bibinfo {author} {\bibfnamefont {B.}~\bibnamefont
  {D\'ora}},\ }\href {\doibase 10.1103/PhysRevLett.115.160402} {\bibfield
  {journal} {\bibinfo  {journal} {Phys. Rev. Lett.}\ }\textbf {\bibinfo
  {volume} {115}},\ \bibinfo {pages} {160402} (\bibinfo {year}
  {2015})}\BibitemShut {NoStop}%
\bibitem [{\citenamefont {Zheng}\ and\ \citenamefont
  {Cooper}(2016)}]{zheng2016superradiance}%
  \BibitemOpen
  \bibfield  {author} {\bibinfo {author} {\bibfnamefont {W.}~\bibnamefont
  {Zheng}}\ and\ \bibinfo {author} {\bibfnamefont {N.~R.}\ \bibnamefont
  {Cooper}},\ }\href {\doibase 10.1103/PhysRevLett.117.175302} {\bibfield
  {journal} {\bibinfo  {journal} {Phys. Rev. Lett.}\ }\textbf {\bibinfo
  {volume} {117}},\ \bibinfo {pages} {175302} (\bibinfo {year}
  {2016})}\BibitemShut {NoStop}%
\bibitem [{\citenamefont {{Ballantine}}\ \emph {et~al.}(2016)\citenamefont
  {{Ballantine}}, \citenamefont {{Lev}},\ and\ \citenamefont
  {{Keeling}}}]{keeling_meissner_cav_2016}%
  \BibitemOpen
  \bibfield  {author} {\bibinfo {author} {\bibfnamefont {K.~E.}\ \bibnamefont
  {{Ballantine}}}, \bibinfo {author} {\bibfnamefont {B.~L.}\ \bibnamefont
  {{Lev}}}, \ and\ \bibinfo {author} {\bibfnamefont {J.}~\bibnamefont
  {{Keeling}}},\ }\href@noop {} {\bibfield  {journal} {\bibinfo  {journal}
  {ArXiv e-prints}\ } (\bibinfo {year} {2016})},\ \Eprint
  {http://arxiv.org/abs/1608.07246} {arXiv:1608.07246 [cond-mat.quant-gas]}
  \BibitemShut {NoStop}%
\bibitem [{\citenamefont {Piazza}\ and\ \citenamefont
  {Ritsch}(2015)}]{piazza_blue}%
  \BibitemOpen
  \bibfield  {author} {\bibinfo {author} {\bibfnamefont {F.}~\bibnamefont
  {Piazza}}\ and\ \bibinfo {author} {\bibfnamefont {H.}~\bibnamefont
  {Ritsch}},\ }\href {\doibase 10.1103/PhysRevLett.115.163601} {\bibfield
  {journal} {\bibinfo  {journal} {Phys. Rev. Lett.}\ }\textbf {\bibinfo
  {volume} {115}},\ \bibinfo {pages} {163601} (\bibinfo {year}
  {2015})}\BibitemShut {NoStop}%
\bibitem [{\citenamefont {Zak}(1989)}]{zak_1989}%
  \BibitemOpen
  \bibfield  {author} {\bibinfo {author} {\bibfnamefont {J.}~\bibnamefont
  {Zak}},\ }\href {\doibase 10.1103/PhysRevLett.62.2747} {\bibfield  {journal}
  {\bibinfo  {journal} {Phys. Rev. Lett.}\ }\textbf {\bibinfo {volume} {62}},\
  \bibinfo {pages} {2747} (\bibinfo {year} {1989})}\BibitemShut {NoStop}%
\bibitem [{\citenamefont {Su}\ \emph {et~al.}(1979)\citenamefont {Su},
  \citenamefont {Schrieffer},\ and\ \citenamefont {Heeger}}]{ssh_1979}%
  \BibitemOpen
  \bibfield  {author} {\bibinfo {author} {\bibfnamefont {W.~P.}\ \bibnamefont
  {Su}}, \bibinfo {author} {\bibfnamefont {J.~R.}\ \bibnamefont {Schrieffer}},
  \ and\ \bibinfo {author} {\bibfnamefont {A.~J.}\ \bibnamefont {Heeger}},\
  }\href {\doibase 10.1103/PhysRevLett.42.1698} {\bibfield  {journal} {\bibinfo
   {journal} {Phys. Rev. Lett.}\ }\textbf {\bibinfo {volume} {42}},\ \bibinfo
  {pages} {1698} (\bibinfo {year} {1979})}\BibitemShut {NoStop}%
\bibitem [{\citenamefont {Holstein}(1959)}]{holstein_1959}%
  \BibitemOpen
  \bibfield  {author} {\bibinfo {author} {\bibfnamefont {T.}~\bibnamefont
  {Holstein}},\ }\href@noop {} {\bibfield  {journal} {\bibinfo  {journal}
  {Annals of physics}\ }\textbf {\bibinfo {volume} {8}},\ \bibinfo {pages}
  {325} (\bibinfo {year} {1959})}\BibitemShut {NoStop}%
\bibitem [{\citenamefont {Hasan}\ and\ \citenamefont
  {Kane}(2010)}]{hasan_kane}%
  \BibitemOpen
  \bibfield  {author} {\bibinfo {author} {\bibfnamefont {M.~Z.}\ \bibnamefont
  {Hasan}}\ and\ \bibinfo {author} {\bibfnamefont {C.~L.}\ \bibnamefont
  {Kane}},\ }\href {\doibase 10.1103/RevModPhys.82.3045} {\bibfield  {journal}
  {\bibinfo  {journal} {Rev. Mod. Phys.}\ }\textbf {\bibinfo {volume} {82}},\
  \bibinfo {pages} {3045} (\bibinfo {year} {2010})}\BibitemShut {NoStop}%
\bibitem [{\citenamefont {Ryu}\ \emph {et~al.}(2010)\citenamefont {Ryu},
  \citenamefont {Schnyder}, \citenamefont {Furusaki},\ and\ \citenamefont
  {Ludwig}}]{ryu2010topological}%
  \BibitemOpen
  \bibfield  {author} {\bibinfo {author} {\bibfnamefont {S.}~\bibnamefont
  {Ryu}}, \bibinfo {author} {\bibfnamefont {A.~P.}\ \bibnamefont {Schnyder}},
  \bibinfo {author} {\bibfnamefont {A.}~\bibnamefont {Furusaki}}, \ and\
  \bibinfo {author} {\bibfnamefont {A.~W.}\ \bibnamefont {Ludwig}},\
  }\href@noop {} {\bibfield  {journal} {\bibinfo  {journal} {New Journal of
  Physics}\ }\textbf {\bibinfo {volume} {12}},\ \bibinfo {pages} {065010}
  (\bibinfo {year} {2010})}\BibitemShut {NoStop}%
\bibitem [{\citenamefont {Baumann}\ \emph {et~al.}(2011)\citenamefont
  {Baumann}, \citenamefont {Mottl}, \citenamefont {Brennecke},\ and\
  \citenamefont {Esslinger}}]{eth_jumps}%
  \BibitemOpen
  \bibfield  {author} {\bibinfo {author} {\bibfnamefont {K.}~\bibnamefont
  {Baumann}}, \bibinfo {author} {\bibfnamefont {R.}~\bibnamefont {Mottl}},
  \bibinfo {author} {\bibfnamefont {F.}~\bibnamefont {Brennecke}}, \ and\
  \bibinfo {author} {\bibfnamefont {T.}~\bibnamefont {Esslinger}},\ }\href
  {\doibase 10.1103/PhysRevLett.107.140402} {\bibfield  {journal} {\bibinfo
  {journal} {Phys. Rev. Lett.}\ }\textbf {\bibinfo {volume} {107}},\ \bibinfo
  {pages} {140402} (\bibinfo {year} {2011})}\BibitemShut {NoStop}%
\bibitem [{\citenamefont {Brennecke}\ \emph {et~al.}(2013)\citenamefont
  {Brennecke}, \citenamefont {Mottl}, \citenamefont {Baumann}, \citenamefont
  {Landig}, \citenamefont {Donner},\ and\ \citenamefont
  {Esslinger}}]{eth_non_eq}%
  \BibitemOpen
  \bibfield  {author} {\bibinfo {author} {\bibfnamefont {F.}~\bibnamefont
  {Brennecke}}, \bibinfo {author} {\bibfnamefont {R.}~\bibnamefont {Mottl}},
  \bibinfo {author} {\bibfnamefont {K.}~\bibnamefont {Baumann}}, \bibinfo
  {author} {\bibfnamefont {R.}~\bibnamefont {Landig}}, \bibinfo {author}
  {\bibfnamefont {T.}~\bibnamefont {Donner}}, \ and\ \bibinfo {author}
  {\bibfnamefont {T.}~\bibnamefont {Esslinger}},\ }\href {\doibase
  10.1073/pnas.1306993110} {\bibfield  {journal} {\bibinfo  {journal}
  {Proceedings of the National Academy of Sciences}\ }\textbf {\bibinfo
  {volume} {110}},\ \bibinfo {pages} {11763} (\bibinfo {year}
  {2013})}\BibitemShut {NoStop}%
\bibitem [{\citenamefont {Klinder}\ \emph {et~al.}(2014)\citenamefont
  {Klinder}, \citenamefont {Ke\ss{}ler}, \citenamefont {Wolke}, \citenamefont
  {Mathey},\ and\ \citenamefont {Hemmerich}}]{hemmerich_dyn_2014}%
  \BibitemOpen
  \bibfield  {author} {\bibinfo {author} {\bibfnamefont {J.}~\bibnamefont
  {Klinder}}, \bibinfo {author} {\bibfnamefont {H.}~\bibnamefont {Ke\ss{}ler}},
  \bibinfo {author} {\bibfnamefont {M.}~\bibnamefont {Wolke}}, \bibinfo
  {author} {\bibfnamefont {L.}~\bibnamefont {Mathey}}, \ and\ \bibinfo {author}
  {\bibfnamefont {A.}~\bibnamefont {Hemmerich}},\ }\href@noop {} {\bibfield
  {journal} {\bibinfo  {journal} {arXiv preprint arXiv:1409.1945}\ } (\bibinfo
  {year} {2014})}\BibitemShut {NoStop}%
\bibitem [{\citenamefont {Resta}(1994)}]{resta_rmp}%
  \BibitemOpen
  \bibfield  {author} {\bibinfo {author} {\bibfnamefont {R.}~\bibnamefont
  {Resta}},\ }\href {\doibase 10.1103/RevModPhys.66.899} {\bibfield  {journal}
  {\bibinfo  {journal} {Rev. Mod. Phys.}\ }\textbf {\bibinfo {volume} {66}},\
  \bibinfo {pages} {899} (\bibinfo {year} {1994})}\BibitemShut {NoStop}%
\bibitem [{\citenamefont {King-Smith}\ and\ \citenamefont
  {Vanderbilt}(1993)}]{vanderbilt_1993}%
  \BibitemOpen
  \bibfield  {author} {\bibinfo {author} {\bibfnamefont {R.~D.}\ \bibnamefont
  {King-Smith}}\ and\ \bibinfo {author} {\bibfnamefont {D.}~\bibnamefont
  {Vanderbilt}},\ }\href {\doibase 10.1103/PhysRevB.47.1651} {\bibfield
  {journal} {\bibinfo  {journal} {Phys. Rev. B}\ }\textbf {\bibinfo {volume}
  {47}},\ \bibinfo {pages} {1651} (\bibinfo {year} {1993})}\BibitemShut
  {NoStop}%
\bibitem [{\citenamefont {Atala}\ \emph {et~al.}(2013)\citenamefont {Atala},
  \citenamefont {Aidelsburger}, \citenamefont {Barreiro}, \citenamefont
  {Abanin}, \citenamefont {Kitagawa}, \citenamefont {Demler},\ and\
  \citenamefont {Bloch}}]{bloch_zak}%
  \BibitemOpen
  \bibfield  {author} {\bibinfo {author} {\bibfnamefont {M.}~\bibnamefont
  {Atala}}, \bibinfo {author} {\bibfnamefont {M.}~\bibnamefont {Aidelsburger}},
  \bibinfo {author} {\bibfnamefont {J.~T.}\ \bibnamefont {Barreiro}}, \bibinfo
  {author} {\bibfnamefont {D.}~\bibnamefont {Abanin}}, \bibinfo {author}
  {\bibfnamefont {T.}~\bibnamefont {Kitagawa}}, \bibinfo {author}
  {\bibfnamefont {E.}~\bibnamefont {Demler}}, \ and\ \bibinfo {author}
  {\bibfnamefont {I.}~\bibnamefont {Bloch}},\ }\href
  {http://dx.doi.org/10.1038/nphys2790} {\bibfield  {journal} {\bibinfo
  {journal} {Nat Phys}\ }\textbf {\bibinfo {volume} {9}},\ \bibinfo {pages}
  {795} (\bibinfo {year} {2013})}\BibitemShut {NoStop}%
\bibitem [{\citenamefont {Peierls}(1955)}]{peierls_1955}%
  \BibitemOpen
  \bibfield  {author} {\bibinfo {author} {\bibfnamefont {R.}~\bibnamefont
  {Peierls}},\ }\href {http://books.google.de/books?id=WvPcBUsSJBAC} {\emph
  {\bibinfo {title} {Quantum Theory of Solids}}},\ International series of
  monographs on physics\ (\bibinfo  {publisher} {Clarendon Press},\ \bibinfo
  {year} {1955})\BibitemShut {NoStop}%
\bibitem [{\citenamefont {Ryu}\ and\ \citenamefont
  {Hatsugai}(2002)}]{hatsugai_2002}%
  \BibitemOpen
  \bibfield  {author} {\bibinfo {author} {\bibfnamefont {S.}~\bibnamefont
  {Ryu}}\ and\ \bibinfo {author} {\bibfnamefont {Y.}~\bibnamefont {Hatsugai}},\
  }\href {\doibase 10.1103/PhysRevLett.89.077002} {\bibfield  {journal}
  {\bibinfo  {journal} {Phys. Rev. Lett.}\ }\textbf {\bibinfo {volume} {89}},\
  \bibinfo {pages} {077002} (\bibinfo {year} {2002})}\BibitemShut {NoStop}%
\bibitem [{Note1()}]{Note1}%
  \BibitemOpen
  \bibinfo {note} {See Supplemental Material}\BibitemShut {NoStop}%
\bibitem [{Note2()}]{Note2}%
  \BibitemOpen
  \bibinfo {note} {The generalization to non equilibrium steady-states in
  presence of cavity losses is done in the Ref.~\cite
  {piazza_QKE}.}\BibitemShut {Stop}%
\end{thebibliography}%

\newpage
\widetext

\section{Derivation of the mean-field equations and the atomic polarization function}

We will now derive the formula for the atomic polarization function $\chi(\omega)$,
i.e., the linear response of the atoms to density perturbations induced
by the cavity photons. This can be done by computing the cavity photon
propagator including the interactions with the atoms at the
random-phase approximation level, where $\chi(\omega)$ appears as
a one-loop self-energy correction.
Within the equilibrium~\footnote{The generalization to non equilibrium
  steady-states in presence of cavity losses is done
  in the Ref.~\cite{piazza_QKE}.} path-integral approach developed in~\cite{piazza_bose,piazza_fermi}, we can integrate out the atomic
degrees of freedom and obtain the effective action for the cavity
photons
\begin{align}
S_{\rm eff}=\sum_{\nu}(-i\omega_\nu-\Delta_c)|a_\nu|^2+\mathrm{Tr}\mathrm{ln}(M),
\label{eq:Seff}
\end{align}
where $a_\nu$ is the cavity field component corresponding to the fermionic
Matsubara frequency $\omega_\nu=k_BT(2\nu+1)\pi$, $\nu\in\mathbb{Z}$
and the trace reads $\mathrm{Tr}=\sum_\nu\int\!\! dx$ with the matrix element
\begin{align}
M_{\nu,\nu'}(x,x')=G_{\nu,\nu'}^{-1}(x,x')+A_{\nu,\nu'}(x,x'),
\label{eq:M}
\end{align}
where
\[
G_{\nu,\nu'}^{-1}(x,x')=\delta_{\nu,\nu'}\left[-i\omega_\nu
 +H_{\rm sp}-\mu\right]\;,\;H_{\rm sp}= -\frac{\partial_{xx}^2}{2m}+V_{\rm sl}(x),
\]
and
\[
A_{\nu,\nu'}^{-1}(x,x')=\delta(x-x')\left[\frac{\partial^2 V_{\rm
    sl}(x)}{\partial\alpha^*\partial\alpha}\sum_{\nu_1}\delta a_{\nu_1}^*\delta a_{\nu_1+\nu-\nu'}+\frac{\partial V_{\rm
    sl}(x)}{\partial\alpha^*}\left(\delta a_{\nu-\nu'}+\delta a_{\nu'-\nu}^*\right)\right],
\]
with 
\[
V_{\rm  sl}(x)=U_0|\alpha|^2\cos^2(k_cx)+2\eta |\alpha|\cos(\Delta\varphi)\cos(k_cx).
\]
Here, we have split the cavity field $a_\nu=\alpha+\delta a_\nu$ in
its coherent component (assumed to be real) and fluctuations.

By expanding the logarithm in \eqref{eq:Seff} up to second order in
the fluctuations $\delta a$ we get 
\begin{align}
S_{\rm eff}\simeq S_{\rm eff}^{\rm
  MF}+\sum_{\nu}\left(\begin{array}{cc}\delta a_\nu^* & \delta a_{-\nu}\end{array}\right)D^{-1}(\omega_\nu)\left(\begin{array}{c}\delta a_\nu \\ \delta a_{-\nu}^*\end{array}\right).
\label{eq:Seff2}
\end{align}
The mean-field action $S_{\rm eff}^{\rm
  MF}=-\Delta_c|\alpha|^2+\mathrm{Tr}\mathrm{ln}(G^{-1})$ contains in particular the
Josephson-like energy term $E_J$ discussed in the main text. The
mean-field equation for $\alpha$ of the main text is obtained from
\[
\frac{\partial S_{\rm eff}^{\rm  MF}}{\partial\alpha^*}=0,
\] 
and using 
\[
\frac{\partial}{\partial\alpha^*}\mathrm{Tr}\mathrm{ln}(G^{-1})=\mathrm{Tr}(G\frac{\partial
G^{-1}}{\partial\alpha^*})=\sum_\nu\int\!\! dx \rho(x)\frac{\partial
  V_{\rm sl}(x)}{\partial\alpha^*}.
\]
In Eq.~\eqref{eq:Seff2} the inverse photon propagator is given by
\[
D^{-1}(\omega_\nu)=\left(\begin{array}{cc}-i\omega_\nu-\Delta_c+\sum_\ell
    n_F(\epsilon_\ell)\int\!\! dx \frac{\partial V_{\rm
    sl}}{\partial\alpha^*}|\psi_\ell|^2 +\chi(\omega_\nu)& \chi(\omega_\nu)\\ \chi(-\omega_\nu)&i\omega_\nu-\Delta_c+\sum_\ell
    n_F(\epsilon_\ell)\int\!\! dx \frac{\partial V_{\rm
    sl}}{\partial\alpha^*}|\psi_\ell|^2 +\chi(-\omega_\nu)\end{array}\right),
\]
and contains the polarization function
\[
\chi(\omega_\nu)=\sum_{\ell,\ell'}\frac{n_F(\epsilon_\ell)-n_F(\epsilon_{\ell'})}{i\omega_\nu+\epsilon_{\ell}
  -\epsilon_{\ell'}}\bigg|\int\!\! dx\psi_\ell^*(x) \frac{\partial V_{\rm
    sl}(x)}{\partial\alpha^*}\psi_{\ell'}(x)\bigg|^2,
\]
which coincides with the expression given in the main text after analytic
continuation $\omega_\nu\to -i\omega+0^+$.

\end{document}